\documentclass[galaxies,article,accept,pdftex,moreauthors]{Definitions/mdpi}

\usepackage{makecell}
\usepackage{bigints}
\usepackage{tikz}
\usepackage[normalem]{ulem}

\firstpage{1} 
\pubvolume{1}
\issuenum{1}
\articlenumber{0}
\pubyear{2023}
\copyrightyear{2023}
\externaleditor{\textls[-15]{}\vspace{-12pt}}
\datereceived{7 December 2023} 
\daterevised{19 January 2024} 
\dateaccepted{20 January 2024} 
\datepublished{ } 
\hreflink{https://doi.org/} 

\Title{The 
Scavenger Hunt for Quasar Samples to Be Used as Cosmological Tools}

\TitleCitation{The Scavenger Hunt for Quasar Samples to Be Used as Cosmological Tools}

\Author{{Maria Giovanna Dainotti} 
$^{1,2,3,4,}${$^{\dagger}$}
\orcidA{}, Giada Bargiacchi $^{5,6,}${*$^{,\dagger}$}\orcidB{}, Aleksander Łukasz Lenart $^{7}$\orcidC{} \linebreak  and Salvatore Capozziello $^{5,6,8}${\orcidD{}} 
}

\AuthorNames{Maria Giovanna Dainotti, Giada Bargiacchi, Aleksander Łukasz Lenart and Salvatore Capozziello}

\AuthorCitation{Dainotti, M.G.; Bargiacchi, G.; Lenart A.Ł.; Capozziello, S.}

\address{%
$^{1}$ \quad National Astronomical Observatory of Japan, 2 Chome-21-1 Osawa, Mitaka, Tokyo 181-8588, Japan; {maria.dainotti@nao.ac.jp} 
\\
$^{2}$ \quad The Graduate {University} 
for Advanced Studies, {SOKENDAI} 
, Shonankokusaimura, Hayama, Miura District, Kanagawa 240-0193, Japan\\
$^{3}$ \quad Space Science Institute, 4765 Walnut St, Suite B, Boulder, CO 80301, USA\\
$^{4}$ \quad Department of Physics and Astrophysics, University of Las Vegas, Las Vegas, NV 89154, USA\\
$^{5}$ \quad Scuola Superiore Meridionale, Largo S. Marcellino 10, 80138 Napoli, Italy; {capozziello@na.infn.it}\\
$^{6}$ \quad Istituto Nazionale di Fisica Nucleare (INFN), Sez. di Napoli, Complesso University Monte S. Angelo, Via Cinthia 9, 80126 Napoli, Italy\\
${^7}$ \quad Astronomical Observatory, Jagiellonian University in Kraków, Orla 171, 30-244 Krakow, Poland; {aleksander.lenart@student.uj.edu.pl}\\
$^{8}$ \quad Dipartimento di Fisica ``E. Pancini'', Università degli Studi di Napoli Federico II, Complesso University Monte S. Angelo, 
Via Cinthia 9, 80126 Napoli, Italy}

\corres{{Correspondence:} 
maria.dainotti@nao.ac.jp}

\firstnote{The first two authors contributed equally to this work.} 

\abstract{Although the $\Lambda$ Cold Dark Matter model is the most accredited cosmological model, information at high redshifts ($z$) between type Ia supernovae ($z=2.26$) and the Cosmic Microwave Background ($z=1100$) is
crucial to validate this model further. To this end, we have discovered a sample of 1132 quasars up to $z=7.54$ exhibiting a reduced intrinsic dispersion of the relation between ultraviolet and X-ray fluxes, $\delta_\mathrm{F}=0.22$ vs. $\delta_\mathrm{F}=0.29$ ($24\%$ less), than the original sample. This gold sample, once we correct the luminosities for selection biases and redshift evolution, enables us to determine the matter density parameter $\Omega_M$ with a precision of 0.09. Unprecedentedly, this quasar sample is the only one that, as a standalone cosmological probe, yields such tight constraints on $\Omega_M$ while being drawn from the same parent population of the initial sample. 
}

\begin{document}



\section{Introduction}
\label{introduction}
Recently, improved precision in measuring cosmological parameters has exposed tantalizing discrepancies within the widely accepted $\Lambda$ Cold Dark Matter (CDM) model. 
This model describes the Universe relying on a CDM and dark energy components, where the dark energy is a cosmological constant ($\Lambda$), as required by the current accelerated expansion of the Universe~\citep{riess1998,perlmutter1999}. 
This model, with its enigmatic dark energy and cold dark matter, has served us well, explaining phenomena like the Cosmic Microwave Background (CMB)~\citep{planck2018} and the accelerated expansion of the Universe proved by type Ia supernovae (SNe Ia). Despite its advantages, theoretical flaws still need to be understood.
This is the case of the cosmological constant problem~\citep{1989RvMP...61....1W}, which is the tension between the predicted and observed values of $\Lambda$, the nature of dark energy and its origin, and the fine-tuning problem, which derives from the fact that the current values of the matter density ($\Omega_M$) and the dark energy density ($ \Omega_{\Lambda}$) are of the same order, whereas this is not expected due to their different evolution in time. In addition to these issues, recent measurements have highlighted the so-called Hubble constant ($H_0$) tension. This is the discrepancy between the value of $H_0$ measured locally from SNe Ia and Cepheids, which is $H_0 = 73.04 \pm 1.04  \, \mathrm{km} \, \mathrm{s}^{-1} \, \mathrm{Mpc}^{-1}$~\citep{2022ApJ...934L...7R}, and the value of $H_0$ extrapolated from the Planck data on the CMB within a flat $\Lambda$CDM model, $H_0 = 67.4 \pm 0.5  \, \mathrm{km} \, \mathrm{s}^{-1} \, \mathrm{Mpc}^{-1}$~\citep{planck2018}. The difference between these two measurements ranges between 4.4 and 6$\sigma$, according to the samples investigated~\citep{2019ApJ...876...85R,2020PhRvR...2a3028C,2020MNRAS.498.1420W}. However, the maximum redshift reached by SNe Ia observations is $z=2.26$~\citep{Rodney}, while the CMB radiation is observed at $z=1100$. Thus, it is crucial to probe the Universe in the intermediate epochs between these two to shed light on this tension, hence to confirm, alleviate, or even solve it. To this end, other probes rather than SNe Ia and CMB have already been investigated. These analyses have provided even a more complicated context: cosmic chronometers show a preference for the $H_0$ value derived from the CMB~\citep{2018JCAP...04..051G}, time delay and strong lensing from Quasars (QSOs) favor the $H_0$ from SNe Ia~\citep{2019ApJ...886L..23L}, while QSOs~\citep{biasfreeQSO2022}, the Tip of the Red-Giant Branch~\citep{2021ApJ...919...16F}, and Gamma-Ray Bursts (GRBs)~\citep{Dainotti2022MNRAS.tmp.2639D,Dainotti2022PASJ} hint at an intermediate value of $H_0$ between the one of the CMB and the $H_0$ of SNe Ia.
For the case of QSOs and GRBs, the value of $H_0$ obtained depends on several factors, such as if they are calibrated or not with SNe Ia, if they are fitted jointly with SNe Ia, and if their luminosities have been corrected for the redshift evolution and selection biases. Moreover, we here note that, when also SNe Ia are used, $H_0$ is a parameter degenerate with the absolute magnitude of SNe Ia. Nevertheless, the values of $H_0$ obtained, when QSOs or GRBs are considered, show a trend toward an intermediate value between the CMB and SNe Ia~\citep{biasfreeQSO2022,Dainotti2022MNRAS.tmp.2639D}.
To solve this intriguing puzzle and test if the flat $\Lambda$CDM model still represents the most suitable description of the Universe, reliable and powerful cosmological probes at redshift between $z=2.26$ and $z=1100$ are required. 
To crack this perplexing puzzle and put the $\Lambda$CDM model to the ultimate test, we need to voyage through uncharted territories: we are embarking on a mission to explore the Universe between the epochs of SNe Ia and the CMB.
To date, the best candidates for this purpose are GRBs and QSOs.

In this framework, QSOs have recently attracted more and more interest in the cosmological community~\citep{2008mss..conf...99R,rl19,2020MNRAS.497..263K,2022MNRAS.515.1795B,2022arXiv221014432W,biasfreeQSO2022,DainottiQSO,2022MNRAS.517.1901L,2022PhRvD.106l3523P,Bargiacchi2023MNRAS.521.3909B}, since they are observed up to $z=7.64$~\citep{2021ApJ...907L...1W}, at redshifts much higher than the maximum redshift of SNe Ia observations, $z=2.26$ \cite{Rodney}. The method to standardize QSOs as cosmological candles is based on the Risaliti--Lusso (RL) relation between the logarithms of the Ultraviolet (UV) luminosity at $2500$ \AA \, ($L_{UV}$) and the X-ray luminosity at 2 KeV ($L_X$). {{The}  
RL relation reads as $\mathrm{log_{10}} L_X = \gamma \, \mathrm{log_{10}} L_{UV} + \beta$, where the slope $\gamma$ and the intercept $\beta$ usually have values of $\gamma {\sim} 0.6$ and $\beta {\sim} 8$.}
This empirical relation has been validated with several QSO samples~\citep{1979ApJ...234L...9T,1981ApJ...245..357Z,1982ApJ...262L..17A,1986ApJ...305...83A,steffen06,just07,2010ApJ...708.1388Y,2010A&A...512A..34L,lr16,2021A&A...655A.109B} and it has been turned into a cosmological tool via a careful selection of the QSO sources aimed at removing observational biases~\citep{rl15,lr16,2017A&A...602A..79L,2019A&A...632A.109N,rl19,lusso2019,2020A&A...642A.150L,2022MNRAS.515.1795B}. 
To achieve this standard set starting from the initial QSO sample, several QSOs have been discarded through the investigation of different features.
This procedure allows the sample to present well-defined properties, not to be hampered by a low signal-to-noise ratio, and not to be severely affected by extinction, UV reddening, and contamination of the host galaxy (see Section~\ref{data}). The relation between UV and X-ray luminosities is also theoretically supported by the most accredited QSO model in which an accretion disk powers the central supermassive black hole converting mass into energy. In this scenario, the UV emission of the accretion disk is then reprocessed in X-rays from an external region of relativistic electrons via the inverse Compton effect. Nevertheless, this mechanism still needs to be fully understood to explain the stability of the X-ray emission, which is not expected since the electrons should cool down falling on the central region. Thus, this stability requires an efficient energy transfer between central and external regions, whose origin is yet to be unveiled~\citep{2017A&A...602A..79L,2023arXiv231216562B}. Moreover, {Ref.} 
~\citep{DainottiQSO} has confirmed the reliability of the RL relation in cosmology proving that this relation is completely intrinsic to the QSO physics and not induced by selection effects or redshift evolution. 
This is a crucial turning point for the reliable application of QSOs as cosmological tools.
{{From now on, the notation ``RL relation'' refers to the relation obtained by the Risaliti--Lusso group, while the notation ``RL correlation'' is used when the relation is applied for fits, in bins, in the flux--flux space, and in the luminosity--luminosity space.}}

In common practice, QSOs are used jointly with other probes since the intrinsic dispersion, $\delta$, of the RL relation ($\delta {\sim} 0.23$ {\citep{rl19,2020A&A...642A.150L}}) 
still limits their power in constraining cosmological parameters, compared to the precision of other probes, such as SNe Ia. {We here stress that we use the term ``intrinsic dispersion'' to refer to the additional fit parameter of the RL relation which is implemented in the fitting likelihood. This allows for a spread around the ideal RL best fit.} For this reason, we here focus on the scavenger hunt of a sub-sample of QSOs, a ``gold'' sample, which presents the optimal compromise between reduced intrinsic dispersion and a sufficient number of sources to be used as a standalone probe and constrain $\Omega_M$ with unprecedented precision in the QSO realm. Similar efforts have already been made in the GRB domain~\citep{dainotti2015b,Dainotti2022ApJS..261...25D,Dainotti2021PASJ...73..970D,Dainotti2022PASJ,Dainotti2022MNRAS.514.1828D,Dainotti2022MNRAS.tmp.2639D}, leading to the definition of the ``Platinum'' GRB sample, which has been used in several cosmological analyses~\citep{Dainotti2022MNRAS.514.1828D,Dainotti2022MNRAS.tmp.2639D}.
Indeed, to choose a standard candle it is crucial to identify either the morphological properties of the lightcurves or the spectral features of the objects investigated. In the case of GRBs, the morphological feature that drives a standard candle is the plateau emission with peculiar characteristics, {{such as not be a steep plateau (with an angle < 41\% and to have at least 5 data points at the beginning of the lightcurve and not to have flares or gaps inside the plateau region)}} {(see, e.g.,~\citep{Dainotti2016ApJ...825L..20D,Dainotti2017ApJ...848...88D, Dainotti2020ApJ...904...97D} for details)}.
Regarding QSOs, the feature stressed here is the fulfillment of the RL relation.  

{We here also stress the reason why the hunt for a gold cosmological sample, not only for QSOs but also for GRBs, has assumed such a relevant role in the cosmological community. Indeed, as anticipated, due to the intrinsic dispersion of the RL relation, QSOs are not able alone to constrain cosmological parameters with precision. As a consequence, once they are combined with other probes, such as SNe Ia, they are not the “leading” probe, which means that the information that dominates is the one from the other probes.
Besides that, QSOs could be extremely important to add information at $z>1$. So, to make QSOs powerful cosmological probes, we need first to determine a QSO sample that is itself capable of constraining cosmological parameters when used alone. Only after that, we can join this gold sample with other probes. In this case, QSOs would contribute to the determination of cosmological parameters by adding their piece of information and changing, confirming, or eventually even improving the constraints given by the other probes.}
{In view of the hunt for a gold QSO sample,} we here present an 
{unprecedented} sample of QSOs reaching the maximum redshifts up to $z = 7.54$, which can be applied as a standalone cosmological probe to constrain $\Omega_M$.
{We here point out that our main aim is to develop a selection procedure that can be reliably applied to select a QSO sample that can be used alone to constrain cosmological parameters. Indeed, our purpose is not to investigate cosmological tensions, as for example the $H_0$ tension, or to test cosmological models alternative to the standard flat $\Lambda$CDM. This is the reason why we employ our final selected QSO sample to fit the specific case of a flat $\Lambda$CDM model with $H_0$ fixed and only $\Omega_M$ free to vary. This analysis shows that this sample constrains $\Omega_M$ with a precision unprecedentedly reached with only QSOs.}

This manuscript is organized as follows. Section~\ref{data} describes the initial QSO sample. Section~\ref{methods} details the correction of luminosities for selection biases and redshift evolution, the selection of the final QSO sub-samples, and the cosmological fitting method. In Section~\ref{results}, we outline our results, and in Section~\ref{conclusions}, we draw conclusions. {Appendix~\ref{sec:comparison} discusses the different binning approaches investigated and Appendix~\ref{The varying evolution} details the Efron and Petrosian~method.}

\section{The Data Sample}
\label{data}
The initial QSO data set for our analysis is the most recent one released for cosmological applications~\citep{2020A&A...642A.150L}. It counts 2421 sources ranging between $z=0.009$ and $z =7.54$~\citep{banados2018} collected from eight different catalogs in the
literature~\citep{2019A&A...632A.109N,salvestrini2019,2019A&A...630A.118V} and archives~\citep{2016yCat..74570110M,2018A&A...613A..51P,2020A&A...641A.136W,2010ApJS..189...37E}, with the addition of a sub-sample of low-redshift QSOs that present UV observations from the International Ultraviolet Explorer and X-ray data from archives. 
To obtain this QSO sample suitable for cosmological analyses, as many as possible observational biases have been carefully inspected and removed~\citep{rl15,lr16,rl19,salvestrini2019,2020A&A...642A.150L}. We here briefly describe the steps of this selection.
First, only measurements with a sufficient signal-to-noise ratio (\mbox{S/N $\geq 1$}) are retained. Then, QSOs that manifest the presence of extinction (i.e., $E(B-V) > 0.1$) are removed to account for UV reddening and contamination of the host galaxy. 
The contribution of absorption in X-ray is also removed by imposing $\Gamma_X + \Delta \Gamma_X \geq 1.7$ and $\Gamma_X \leq 2.8 $ if $z < 4$ and $\Gamma_X \geq 1.7$ if $z \geq 4$, with $\Gamma_X$ and $\Delta \Gamma_X$ being the photon index and its uncertainty, {where the photon index is the coefficient of the power-law that describes the spectrum in X-ray}. Eventually, the final sample is corrected for the Malmquist bias effect. {Indeed, the Malmquist bias effect states that only larger fluxes can be observed at larger distances; thus, low fluxes are prevented from being seen due to the detector flux limitations, and this creates incomplete samples.} {This effect is overcome by} requiring $\text{log}F_{X,exp} - \text{log}F_{min} \geq {\cal F}$, where $F_{X,exp}$ is the X-ray flux computed from the flux in UV by imposing the RL relation and assuming $\Omega_{M}=0.3$ and $H_{0} = 70\, \mathrm{km\,s^{-1}\,Mpc^{-1}}$ in a flat $\Lambda${CDM} model. {Even though this correction requires the assumption of a specific cosmological model, simulations and mock samples of QSOs have been employed to prove that results are not affected by this choice, as explained in~\citep{2020A&A...642A.150L}.} $F_{min}$ is the minimum observable flux computed for each source from the time of observation of the charge-coupled device~\citep{2001A&A...365L..51W,lr16}. ${\cal F}$ is the threshold value, which is fixed to ${\cal F} = 0.9$ for QSOs derived from the cross-match of the Sloan Digital Sky Survey Data Release 14 (SDSS DR 14) with 4XMM Newton or with XXL, and to ${\cal F} = 0.5$ for the ones with measurements from the SDSS DR 14 and Chandra. To reduce the effects of the X-ray variability, if a source has more than one X-ray observation after this selection, these observations are averaged.
We here notice that the sources that have multiple X-ray observations in the QSO sample of~\citep{2020A&A...642A.150L} are only 289~\citep{2023arXiv231208448S}, which is 12\% of the total sample. Thus, for the remaining 88\% of the sample, we cannot reduce the effect of the variability on the dispersion of the RL relation if we consider that the mechanisms responsible for this variability are still not completely understood. Nevertheless, it is possible to estimate the contribution of the X-ray variability to the observed scatter, as reported in~\citep{2023arXiv231208448S} and discussed in Section~\ref{sec:physicalinterpretation}.
{In this analysis we rely on a selection of a sample already studied in the literature and how the impact of these sources can affect the whole analysis is an interesting subject, but it goes outside the scope of the current paper.}
In our work, we start from this final sample of 2421 sources without any additional selection, such as the cut at redshift $z=0.7$ previously used in some works~\citep{2022MNRAS.515.1795B}, to avoid any possible induced bias due to the reduction in the redshift of the sample~\citep{DainottiQSO,biasfreeQSO2022}.

\section{Methods}
\label{methods}
{Since the 2421 QSOs sample is not yet ready for cosmological use, we here detail a method to find a golden sample, the best, optimal sample useful for cosmological studies.}

\subsection{Selection of the QSO Final Samples}
\label{sec:selection}

To transform QSOs into powerful cosmological probes, we meticulously outline the procedure employed to define our final QSO sub-samples.
The steps of this procedure are detailed below and are justified by physical and theoretical requirements. Indeed, these are necessary steps to define a suitable technique that can be reliably applied to slim the QSO sample aiming at turning them into standalone cosmological probes. To clearly detail and visually show our logical flow, we present the graphical representation of the following steps, regarding our selection methodology in Figure~\ref{fig:algorithm}.

\begin{figure}[H]
\begin{adjustwidth}{-\extralength}{0cm}
\centering
\begin{tikzpicture}[node distance={17mm}, thick, main/.style = {draw, rectangle, align=center, text width=4cm}] 
\node[main] (-1) [text width=4cm, fill=blue!20] {Initial total sample of 2421 sources}; 
\node[main] (0) [below of=-1, text width=4cm, node distance=2.5cm] {Number of bins is determined by the assumed redshift interval. The minimum number in the bin is min(N)=10}; 
\node[main] (05) [below of=0, text width=4cm, node distance=2.5cm] {Choose the redshift condition}; 
\node[main] (051) [below left of=05, text width=3cm, node distance=2.5cm,fill=yellow!30] {$\log_{10}(1/(1+\Delta z))$\\ $ = const.$}; 
\node[main] (052) [below right of=05, text width=3cm, node distance=2.5cm,fill=cyan] {Bins wider in z}; 
\node[main] (053) [right of=052, text width=3cm, node distance=3.5cm,fill=orange!30] {Bins optimized in width}; 
\node[main] (054) [left of=051, text width=3cm, node distance=3.5cm,fill=violet!80] {Bins centered on each QSO}; 
\node[main] (-2) [right of=-1, text width=7.1cm, node distance=6cm,fill=gray!30] {{\bf Definitions}\\ \begin{itemize}
\item $ \delta_{D_{\rm L}}=\log_{10} (D_{\rm L,\,max})- \log_{10} (D_{\rm L,\,min})$.\\ \item $\delta_{int}$---intrinsic scatter of the bin.\\ \item {\bf AD(bin)}---the Anderson--Darling test between the normalized residuals and the normal distribution.\\ \item {\bf The `untouched' sample}---the sample of sources for which it was impossible to create a bin.
\end{itemize}}; 
\node[main] (1) [below of=05, text width=4cm, node distance=3.8cm] {bin = sources fulfilling a given condition}; 
\node[main] (2) [below of=1, text width=4cm, node distance=1.7cm] {Fit ($F_{X}-F_{UV}$); Obtain $\delta_{D_{L}}$, $\delta_{int}$, AD(bin)}; 
\node[main] (3) [below of=2, text width=5cm, node distance=1.5cm] {Is $\delta_{D_{L}}\leq\delta_{int}$ $and \, AD>0.05$?}; 
\node[main] (4) [below left of=3, text width=0.75cm,fill=green] {Yes}; 
\node[main] (5) [below right of=3, text width=0.75cm,fill=red] {No}; 
\node[main] (6) [below left of=4, text width=4cm, node distance=2cm] {We accept the bin as a final}; 
\node[main] (7) [below right of=5, text width=4cm, node distance=2cm] {We move the whole bin to the untouched set}; 
\node[main] (8) [below of=6, text width=4cm, node distance=2cm] {We use the Huber regressor to obtain outlier free bin.}; 
\node[main] (9) [below of=8, text width=4cm, node distance=2cm] {We combine all bins creating our final sample ``without source''.};
\node[main] (10) [right of=9, text width=4cm, node distance=5.3cm, fill=blue!20] {We combine the sample ``without sources'' with the ``untouched set'' creating the sample ``with sources''.};

\draw[->] (-1) -- (0); 
\draw[->] (0) -- (05);
\draw[->] (05) -- (051);
\draw[->] (05) -- (052);
\draw[->] (05) -- (053);
\draw[->] (05) -- (054);
\draw[->] (051) -- (1);
\draw[->] (052) -- (1);
\draw[->] (053) -- (1);
\draw[->] (054) -- (1);
\draw[->] (1) -- (2); 
\draw[->] (2) -- (3); 
\draw[->] (3) -- (4); 
\draw[->] (3) -- (5); 
\draw[->] (4) -- (6); 
\draw[->] (5) -- (7); 
\draw[->] (7) -- (10); 
\draw[->] (6) -- (8); 
\draw[->] (8) -- (9); 
\draw[->] (9) -- (10);

\end{tikzpicture} 
\end{adjustwidth}
\caption{{The} 
flowchart of our selection procedure. The two violet boxes show the start and the end of the scheme, while purple, yellow, blue, and orange boxes identify different binning methods.}
\label{fig:algorithm}
\end{figure}
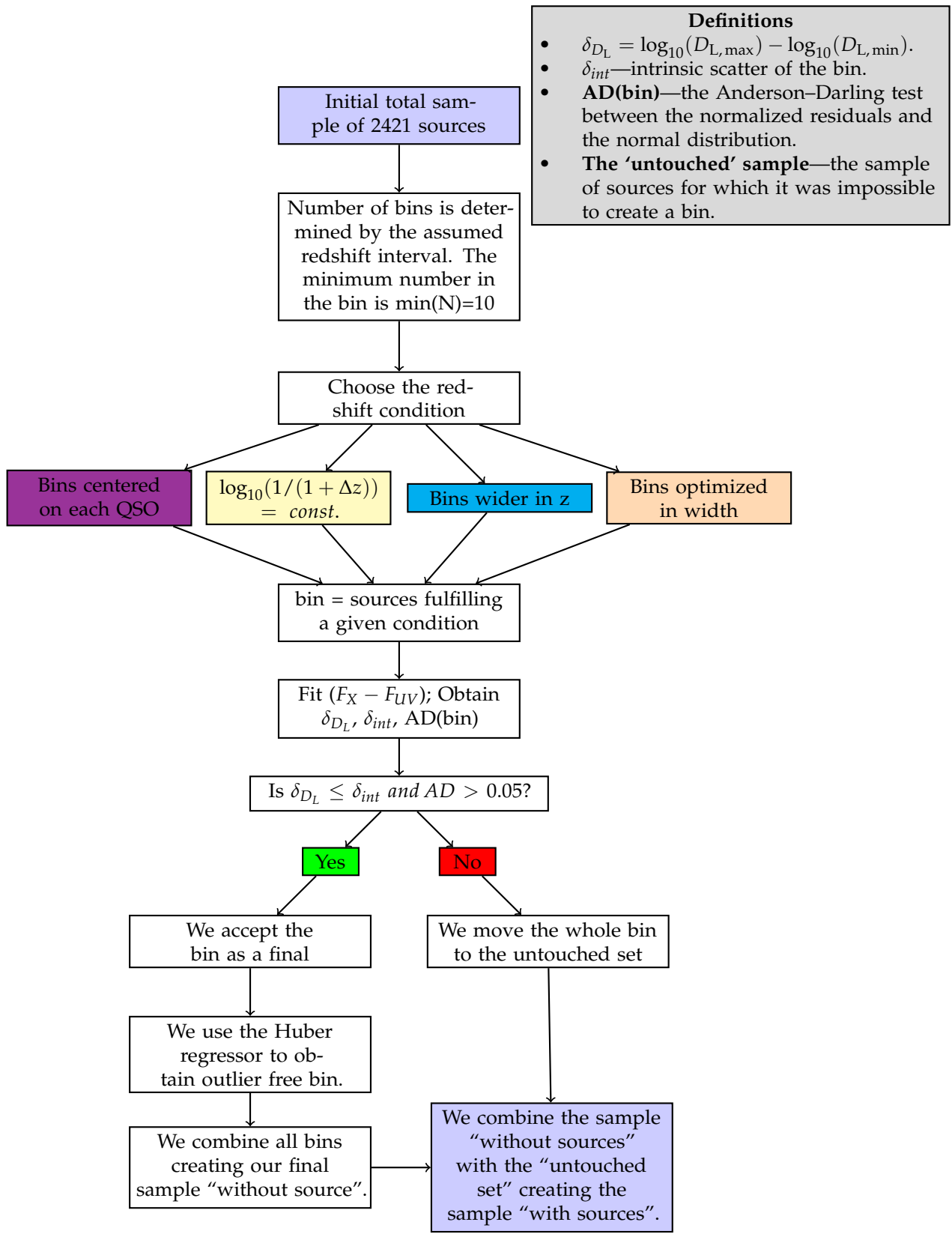
\begin{figure}[H]
\includegraphics[width=12cm]{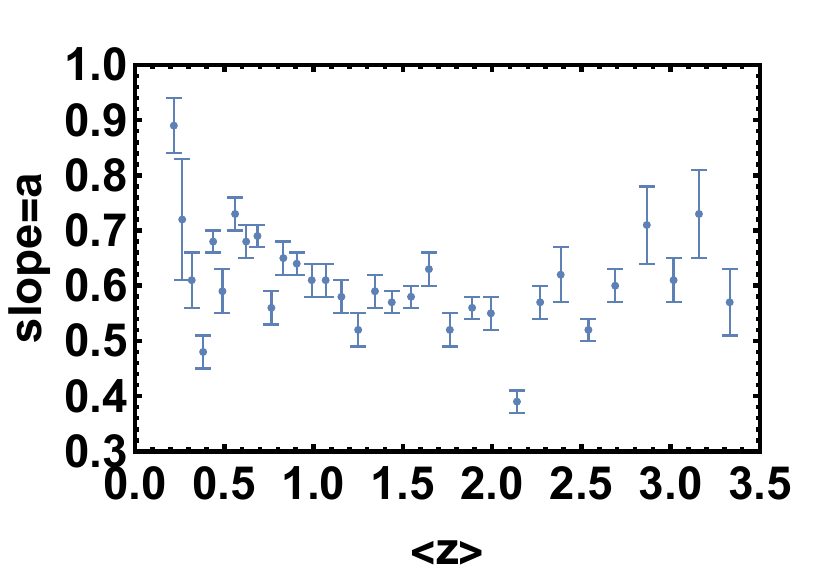}
\caption{{The} 
slope, a, as a function of the average redshift for all bins.}
\label{fig:binvsslope}
\end{figure}

\begin{enumerate}

\item We have first divided the initial QSO sample into bins of redshift to fit a linear relation between the logarithms of fluxes in each redshift bin. The binning must be chosen to verify a specific condition that can be derived from the RL relation~\citep{rl15,rl19,2020A&A...642A.150L,2021A&A...655A.109B}. The RL relation, $\mathrm{log_{10}}L_{X} = \gamma \, \mathrm{log_{10}}L_{UV} + \beta$, can be written in terms of fluxes as
\begin{equation}
\label{RLflux}
\mathrm{log_{10}}F_{X} = a \, \mathrm{log_{10}}F_{UV} + \beta + (a -1) \, \mathrm{log_{10}}(4 \, \pi) + 2 \, (a -1) \, \mathrm{log_{10}}D_L
\end{equation}
where {$a$ is the notation we use to refer to the slope in the flux--flux plane, and} $D_L$ is the luminosity distance in units of cm. Equation~\eqref{RLflux} can be approximated as a linear relation between $\mathrm{log_{10}}F_{X}$ and $\mathrm{log_{10}}F_{UV}$ only if the contribution of $\mathrm{log_{10}}D_L$ is negligible compared to the other terms in Equation~\eqref{RLflux}. Thus, in this case, the linear relation in fluxes represents a proxy of the RL relation in luminosities, with a different intercept, in the form $\mathrm{log_{10}}F_{X} = a \, \mathrm{log_{10}}F_{UV} + b$ and with intrinsic dispersion $\delta_F$. More specifically, if we consider Equation~\eqref{RLflux} in a redshift bin, the contribution of the distance is negligible if the range of values of $\mathrm{log_{10}}D_L$ within the chosen redshift interval is smaller than the intrinsic dispersion of the
relation in the same bin. This is the condition that must be fulfilled when choosing how to divide the QSO sample into redshift bins. Nevertheless, in addition to this requirement, we need to fine-tune our choice to ensure enough sources (we require a minimum of 10), at least in the majority of bins, to reliably perform the fit in each of them. The specific choice of 10 sources is arbitrary. Indeed, we could apply a different threshold, which allows sufficient statistics to perform the fit. We have actually performed our analysis also changing the minimum number to 4, 5, 6, and 10, without any change in our results. 
We here also notice that to reduce the scatter of the relation in each bin, the condition of narrow redshift bins is very relevant, as the smaller the bins, the smaller the difference in $\mathrm{log_{10}}D_L$ is. Hence, we can reduce the intrinsic dispersion up to the limit imposed by $\mathrm{log_{10}}D_L$.
Following all these prescriptions, we have defined our optimal division into redshift bins in terms of $\mathrm{log_{10}}(1/(1+\Delta z))$ with $\Delta z = 0.042$ (see the yellow box in Figure~\ref{fig:algorithm}). We have adopted the division as $\mathrm{log_{10}} (1/(1+z))$, which is a natural choice for the division into redshift since it would retain the same division in volume. This way, we can keep the bin constant, and we do not need to derive arbitrary bins. This is an improvement of the method of bin division in~\citep{2020A&A...642A.150L}.
With this division in redshift, we obtain 32 bins with at least 10 sources (see Table~\ref{tab:bins}), which is the threshold we require to guarantee sufficient statistics for the fit.

We hereby stress that binning the data into redshift intervals is necessary to use fluxes instead of luminosities. This is a crucial point as it enables us to perform a circularity-free analysis. Indeed, fluxes are measured quantities that do not require any cosmological assumption, unlike luminosities. As a consequence, the use of fluxes in the selection of the sample guarantees that our cosmological results are not induced by any a priori cosmological assumption.
It is true that binning leads to the reduction of the sample size (in each bin compared to the total sample size), and therefore, the estimates in each bin might be less accurate. However, in our case, the binning shows that the slope of the flux--flux RL {correlation in each bin} remains unchanged {(see Figure~\ref{fig:binvsslope})}, and is compatible with the slope reported in~\citep{2020A&A...642A.150L}. In addition, binning is often used when it is necessary to highlight features that would otherwise be concealed when noisy data are combined altogether. In this analysis, the binning of the  adopted to avoid the circularity problem (see also~\citep{2023MNRAS.522.1247K} for a discussion on the importance and reliability of the binning method). 
This is because the approximate RL {correlation} for fluxes, which does not depend on the cosmological parameters, holds only within bins of a limited length of redshift and hence of the distance luminosity. This relation within each bin allows us to highlight which QSO sources should be removed. Moreover, we have detailed in Section~\ref{conclusions} that our analysis with the binning gives compatible results with the unbinned data ({see}
~\citep{DainottiGoldQSO2023} for comparison). We also further investigate different choices for the division into bins of the initial sample in Sections~\ref{sec:widerbins}--\ref{sec:nn} and their impact on the cosmological results in {Appendix}
~\ref{sec:comparison}. 
\begin{table}[H]
\label{tab1}
\caption{{Results} 
of the selection procedure in redshift bins {with our fiducial binning in $\mathrm{log_{10}}(1/(1+\Delta z))$ (see yellow box in Figure~\ref{fig:algorithm})}. Mean redshift $<z>$ of each redshift bin with at least 10 sources along with the number of sources N, and the best-fit values of the slope $a$ and the intrinsic dispersion $\delta_F$ with their 1$\sigma$ uncertainty after the removal of outliers.}
\begin{tabularx}{\linewidth}{C C C C} 
\toprule
\textbf{$\boldsymbol{<z>}$} & \textbf{N} & \textbf{$\boldsymbol{a \pm \Delta a}$} & \textbf{$\boldsymbol{\delta_F \pm \Delta \delta_F}$} \\ 
\midrule
0.218 & 7 & $0.89 \pm 0.05$ & $0.06 \pm 0.03$ \\ 
\midrule
0.265 & 9 & $0.72 \pm 0.11$ & $0.11 \pm 0.04$ \\
\midrule
0.318 & 12 & $0.61 \pm 0.05$ & $0.08 \pm 0.02$ \\
\midrule
0.382 & 13 & $0.48 \pm 0.03$ & $0.05 \pm 0.01$ \\
\midrule
0.438 & 13 & $0.68 \pm 0.02$ & $0.01 \pm 0.01$ \\
\midrule
0.490 & 24 & $0.59 \pm 0.04$ & $0.09 \pm 0.02$ \\
\midrule
0.561 & 33 & $0.73 \pm 0.03$ & $0.08 \pm 0.01$ \\
\midrule
0.623 & 33 & $0.68 \pm 0.03$ & $0.06 \pm 0.01$ \\
\midrule
0.686 & 39 & $0.69 \pm 0.02$ & $0.06 \pm 0.01$ \\
\midrule
0.764 & 46 & $0.56 \pm 0.03$ & $0.07 \pm 0.01$ \\
\midrule
0.831 & 48 & $0.65 \pm 0.03$ & $0.07 \pm 0.01$ \\
\midrule
0.906 & 48 & $0.64 \pm 0.02$ & $0.06 \pm 0.01$ \\
\midrule
0.990 & 55 & $0.61 \pm 0.03$ & $0.10 \pm 0.01$ \\
\midrule
1.068 & 44 & $0.61 \pm 0.03$ & $0.06 \pm 0.01$ \\
\midrule
1.156 & 64 & $0.58 \pm 0.03$ & $0.08 \pm 0.01$ \\
\midrule
1.250 & 46 & $0.52 \pm 0.03$ & $0.08 \pm 0.01$ \\
\midrule
1.344 & 48 & $0.59 \pm 0.03$ & $0.08 \pm 0.01$ \\
\midrule
1.438 & 46 & $0.57 \pm 0.02$ & $0.05 \pm 0.01$ \\
\midrule
1.546 & 58 & $0.58 \pm 0.02$ & $0.05 \pm 0.01$ \\
\midrule
1.646 & 49 & $0.63 \pm 0.03$ & $0.06 \pm 0.01$ \\
\midrule
1.763 & 46 & $0.52 \pm 0.03$ & $0.06 \pm 0.01$ \\
\midrule
1.888 & 46 & $0.56 \pm 0.02$ & $0.05 \pm 0.01$ \\
\midrule
1.993 & 45 & $0.55 \pm 0.03$ & $0.06 \pm 0.01$ \\
\midrule
2.139 & 36 & $0.39 \pm 0.02$ & $0.06 \pm 0.01$ \\
\midrule
2.269 & 35 & $0.57 \pm 0.03$ & $0.002 \pm 0.018$ \\
\midrule
2.384 & 32 & $0.62 \pm 0.05$ & $0.03 \pm 0.02$ \\
\midrule
2.539 & 27 & $0.52 \pm 0.02$ & $0.002 \pm 0.014$ \\  
\midrule
2.689 & 22 & $0.60 \pm 0.03$ & $0.04 \pm 0.03$ \\
\midrule
2.866 & 12 & $0.71 \pm 0.07$ & $0.001 \pm 0.012$ \\
\midrule
3.016 & 12 & $0.61 \pm 0.04$ & $0.03 \pm 0.03$ \\
\midrule
3.158 & 12 & $0.73 \pm 0.08$ & $0.05 \pm 0.03$ \\
\midrule
3.330 & 5 & $0.57 \pm 0.06$ & $0.002 \pm 0.019$ \\
\bottomrule
\end{tabularx}
\label{tab:bins}
\end{table}

\item Once we have divided the redshift bins, we fit in each bin that presents at least 10 sources a linear relation between $\mathrm{log_{10}}F_{X}$ and $\mathrm{log_{10}}F_{UV}$. This fit is performed using the Kelly method~\citep{Kelly2007}, which accounts for the uncertainties in both quantities and also for the intrinsic dispersion of the correlation. We have also imposed uniform priors in a wide range of values for the free parameters of the fit: the slope, the intercept, and the intrinsic dispersion. 
To verify that the condition described at Point (1) is satisfied, the best-fit value obtained with the Kelly method for the intrinsic dispersion is compared to the maximum difference of $\mathrm{log_{10}}D_L$ for the sources in the investigated redshift bin. This difference is computed by assuming a flat $\Lambda$CDM model. We here notice that the assumption of a specific cosmological model for this computation does not affect the result since we are considering a difference between two luminosity distances. 
We have retained unmodified sources in the redshift bins that do not provide enough statistics (less than 10 QSOs) to perform a reliable fit. From now on, we denote these sources with the notation ``untouched''.  Also, we have distinguished two cases, one in which we do not include these sources and another one in which we have added them to the final selected sample obtained after Point (3). 
The strategy here is to balance and compromise among the smallest bin so that $\Delta\mathrm{log_{10}}{D_L}< \delta$, but still sufficiently large so that the number of sources is at least 10 or more.

\item At this stage, as the presence of outliers can decrease the performance and accuracy of least-squared-loss error-based regression, we have employed the consolidated statistical technique of the Huber algorithm~\citep{huber1992robust,owen2007robust,ronchetti2009robust} to reduce the intrinsic dispersion in each bin considered.
The Huber regressor is indeed a method for estimating the parameters of a model, in this case the $F_X-F_{UV}$ relation, to detect the outliers and weigh them less in the evaluation of the best-fit parameters of the fitted model.
We are indeed aware that sources more scattered around this relation hamper significantly the finding of the most suitable sample with the smallest intrinsic dispersion. 
Thus, compared to traditional fitting procedures, such as the D’Agostini~\citep{2005physics..11182D} or the Kelly ~\citep{Kelly2007} methods, the Huber regression identifies outliers, which can be caused, for example, by errors or problems in the measurements, and recognizes the actual best-fit based on the inliers.
For these reasons, this technique is widely applied for robust regression problems. 
The Huber regressor has the advantage of not being heavily influenced by the outliers, while not completely ignoring them. This allows us to estimate the actual slope and intercept of the relation, not altered by outliers, and contemporaneously to identify the sources that are outliers of the model. Hence, we discard these sources from the QSO sample in each redshift bin.
In order to quantitatively evaluate the Huber algorithm's numerical gain against the traditional fitting one, we have also compared the results obtained with the Huber regressor with those derived from the traditional sigma-clipping selection technique. This comparison is detailed in Section~\ref{sec:hubergain}.

After this selection, we have also checked in each bin the following criteria: the {null hypothesis that the} populations of both UV and X-ray fluxes {are} drawn from the initial ones in the bin considered {must not be rejected with \emph{p}-value > 5\%} according to the Anderson two-sample test and if the distribution of the residuals about the best-fit line is Gaussian according to the Anderson--Darling normality test with an acceptance significance level of 5\% (see, e.g.,~\citep{2021MNRAS.507..919L} for the Gaussianity discussion).
The Anderson--Darling test for normality determines whether a data sample is drawn from the Gaussian distribution, and it is commonly applied in the literature (e.g.,~\citep{Dainotti2022ApJS..261...25D} in astrophysics and~\citep{stephens1974edf, razali2011power} in statistics). An important property of this test is that it can identify any small deviation from normality. We refer to~\citep{snelikelihood2024} for a detailed description of the features of this test and its application to cosmological likelihoods.
The Anderson--Darling two-sample test instead allows us to verify if the selected sample is still drawn from the original one. This guarantees that we are neither introducing biases nor significantly changing the physical properties of the initial sample when selecting the final sample.
We here also stress that the Anderson two-sample test is always fulfilled at a statistical level >25\%. Table~\ref{tab:bins} reports the mean value of $z$ ($<z>$) for each redshift bin with at least 10 initial sources, 
the number of sources retained, and the corresponding best-fit values for the slope and the intrinsic scatter of the linear relation. A visual representation of the trend of the best-fit values of the slope with the average redshift of each bin is also provided in Figure~\ref{fig:binvsslope}. To showcase the Huber regressor's {advantage} in each bin and how effectively it removes the outliers, in Figure~\ref{fig:bin}, we present in green the selected sample and in red the sources identified as outliers. The two bins investigated on the left and right panels of this figure are the second most populated one and the second least populated one, respectively.

\item As anticipated, we have finally defined two ultimate samples: one with only the sources retained through the steps detailed above and another obtained by combining the sources retained in each bin with the unmodified sources of the bins without enough statistics. This way, we have generated the final selected QSO samples composed of 1065 and 1132 sources, respectively. 
{We here anticipate that, among all the binning approaches investigated in this work, we choose as the best one the one that leads to the best precision on $\Omega_M$ for both samples, with and without untouched sources. The sample obtained with this best method is the one referred to as the ``gold~sample''.}
\end{enumerate}

\vspace{-25pt}
\vspace{-20pt}

\begin{figure}[H]
\begin{adjustwidth}{-\extralength}{0cm}
\centering
\includegraphics[width=.45\linewidth]{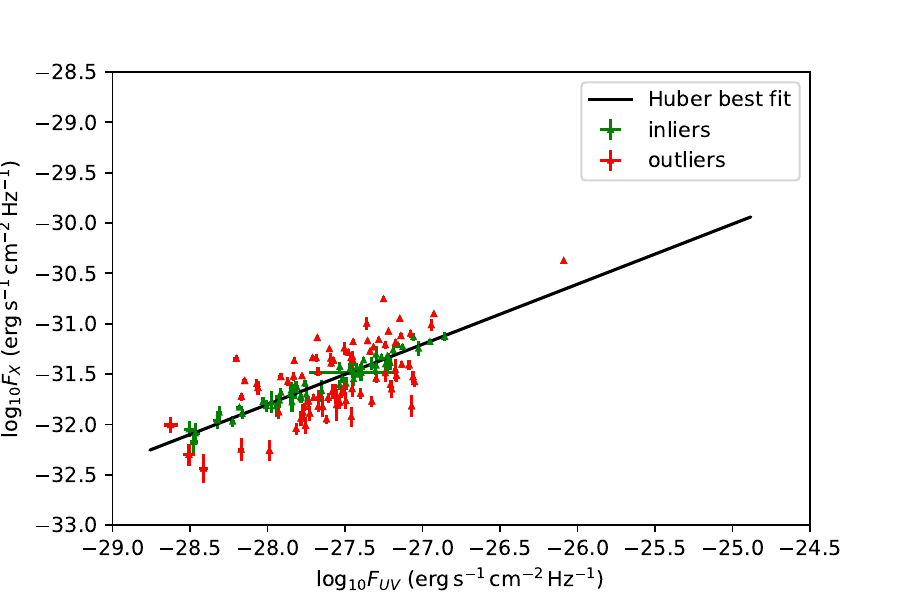}
\includegraphics[width=.45\linewidth]{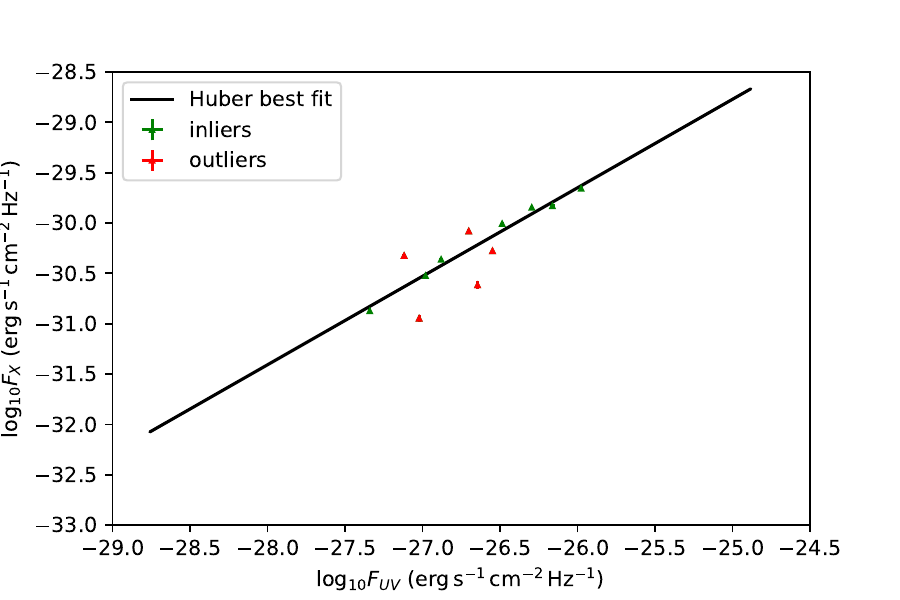}
\end{adjustwidth}
\caption{{Results} 
of the Huber regression in redshift bins. Left panel: Outliers  (in red) and inliers (in green) sources for the second most populated redshift bin along with the Huber best-fit shown with the black line. Right panel: Same as the left panel showing the second least populated redshift bin.}
\label{fig:bin}

\end{figure}
\begin{figure}[H]
\begin{adjustwidth}{-\extralength}{0cm}
\centering
\includegraphics[width=.45\linewidth]{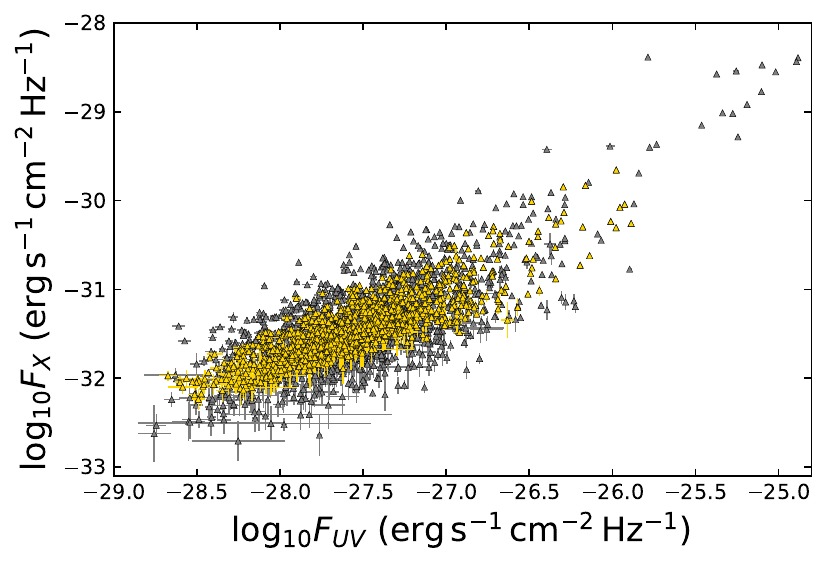} \includegraphics[width=.45\linewidth]{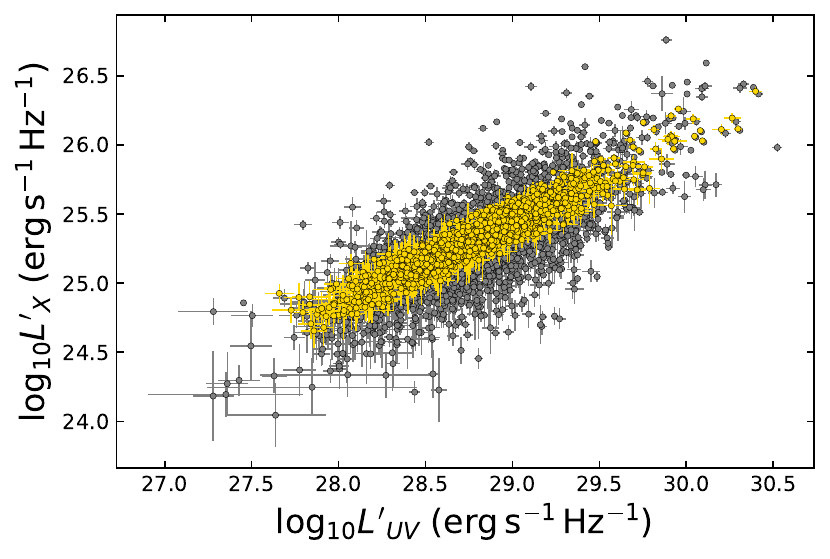}
\end{adjustwidth}

\caption{\textbf{The left panel}: {The} 
QSO sample of 1065 sources (in bright yellow) generated from the $F_\mathrm{X} - F_{\mathrm{UV}}$ relation without adding the sources in redshift bins with insufficient statistics with the best-fit parameters being $a = 0.704 \pm 0.013 $, $b = -12.01 \pm 0.36$, and $\delta_\mathrm{F} = 0.183 \pm 0.004$. The original parent sample is superimposed with gray points. \textbf{Right panel}: The same as the left panel but in the luminosity--luminosity space, once the correction for redshift evolution and selection biases has been applied. In this case, we assume a flat $\Lambda$CDM model with $\Omega_M=0.3$ and $H_0= 70 \, \mathrm{km} \, \mathrm{s}^{-1} \, \mathrm{Mpc}^{-1}$.} 

\label{fig:fxfuv_noadd}
\end{figure}
\begin{figure}[H]
\begin{adjustwidth}{-\extralength}{0cm}
\centering
\includegraphics[width=.45\linewidth]{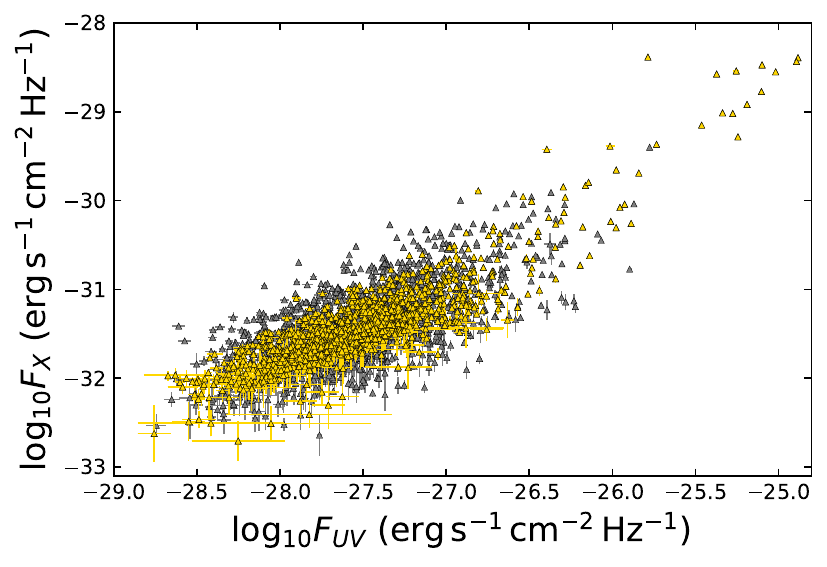} \includegraphics[width=.45\linewidth]{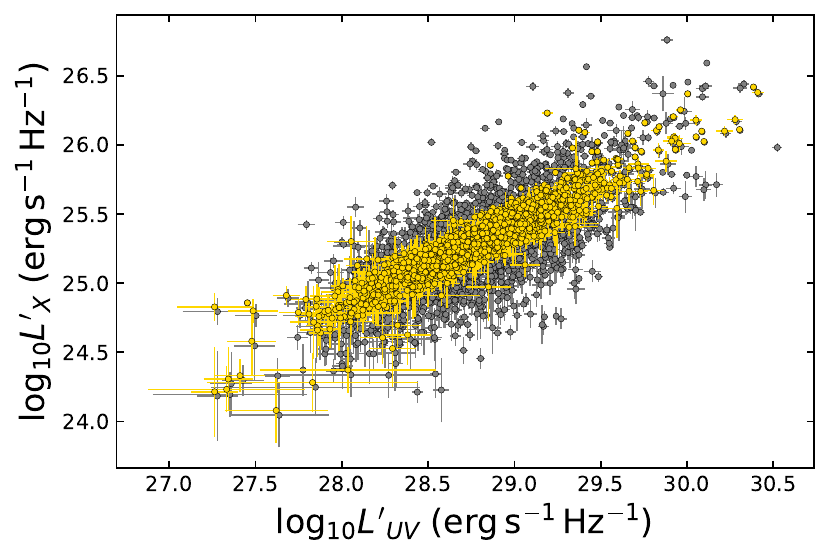}
\end{adjustwidth}
\caption{\textbf{Left panel}: {The} 
flux--flux space with the gold sample of 1132 QSOs (in bright yellow) with best-fit parameters of the linear flux relation: $a = 0.845 \pm 0.013 $, $b = -8.14 \pm 0.36$, and $\delta_\mathrm{F} = 0.223 \pm 0.005$. The original parent sample is superimposed with gray points. The error bars represent the statistical 1$\sigma$ uncertainties. \textbf{Right panel:} Same as the left panel but in the luminosity--luminosity space, once the correction for redshift evolution and selection biases has been applied. In this case, we assume a flat $\Lambda$CDM model with $\Omega_M=0.3$ and $H_0= 70 \, \mathrm{km} \, \mathrm{s}^{-1} \, \mathrm{Mpc}^{-1}$.} 
\label{fig:fxfuv}
\end{figure}
\begin{figure}[H]
\begin{adjustwidth}{-\extralength}{0cm}
\centering
\includegraphics[width=.45\linewidth]{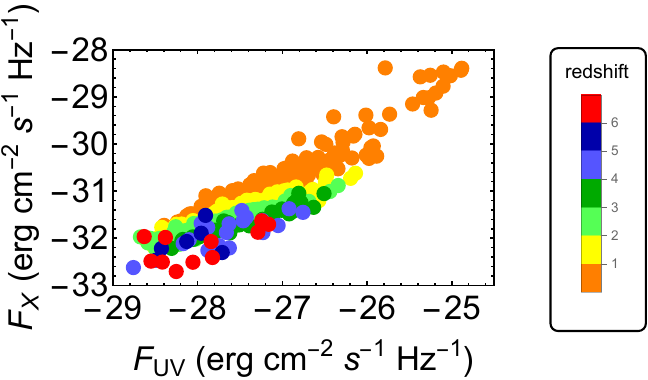} \includegraphics[width=.45\linewidth]{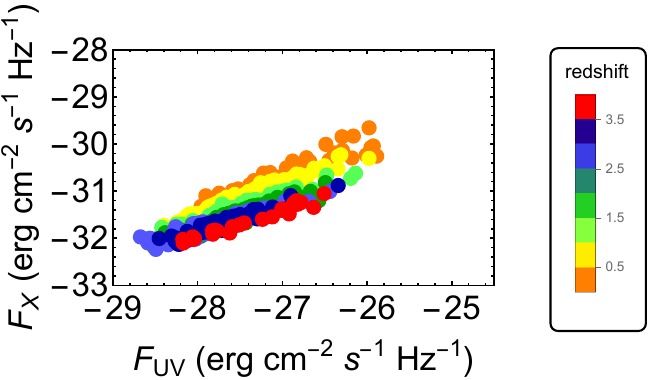}
\end{adjustwidth}
\caption{\textbf{Left panel}:{The} 
logarithmic flux--flux space with the gold sample of 1132 QSOs color coded according to the redshift. \textbf{Right panel}: Same as the left panel but for the smaller sample of 1065 QSOs. }
\label{fig:fxfuvcolorbar}
\end{figure}
\begin{figure}[H]
\begin{adjustwidth}{-\extralength}{0cm}
\centering
\includegraphics[width=.45\linewidth]{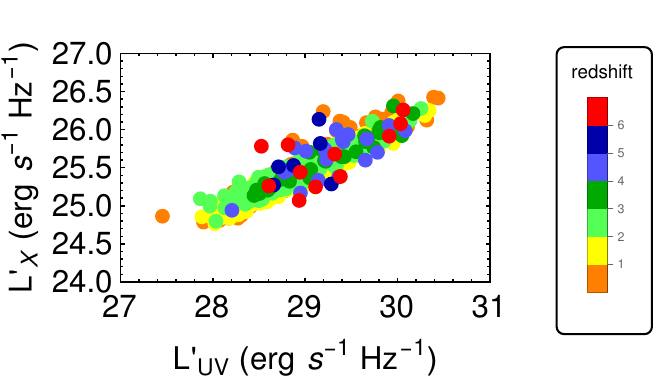} \includegraphics[width=.45\linewidth]{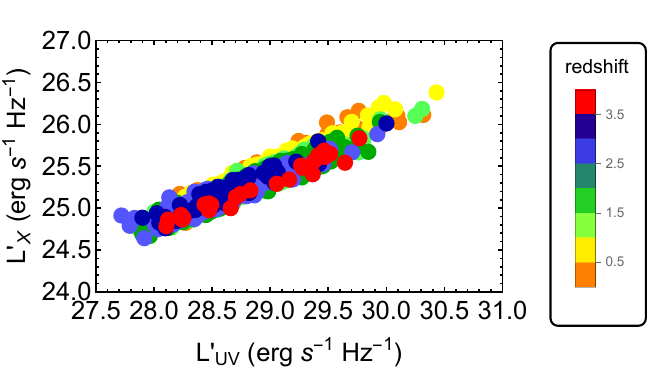}
\end{adjustwidth}
\caption{\textbf{Left panel}: {The} 
logarithmic luminosity--luminosity space corrected for the evolution on both luminosities with the gold sample of 1132 QSOs color coded according to the redshift.  In this case, we assume a flat $\Lambda$CDM model with $\Omega_M=0.3$ and $H_0= 70 \, \mathrm{km} \, \mathrm{s}^{-1} \, \mathrm{Mpc}^{-1}$. \textbf{Right panel}: Same as the left panel but for the smaller sample of 1065 QSOs. }
\label{fig:lxluvcolorbar}
\end{figure}

By using these final QSO samples, we have also fitted the linear relation in fluxes, as shown in the left panels of Figures ~\ref{fig:fxfuv_noadd} and~\ref{fig:fxfuv}. 
As a further step, we have transformed the relation from fluxes to luminosities (see Figures~\ref{fig:fxfuvcolorbar} and~\ref{fig:lxluvcolorbar}) to check if, for both samples, the slope of the luminosity--luminosity {correlation obtained from our analysis} is consistent with the slope of the RL relation corrected for the redshift evolution.
More specifically, we have computed from the fluxes the corresponding luminosities, and we have fitted a linear relation among them with the following form:
\begin{equation}
\mathrm{log_{10}}L'_{X} = \gamma' \, \mathrm{log_{10}}L'_{UV} + \beta'.
\label{RL}
\end{equation}
\vspace{-20pt}

\vspace{-12pt}

\vspace{-20pt}
\begin{figure}[H]
\includegraphics[width=12 cm]{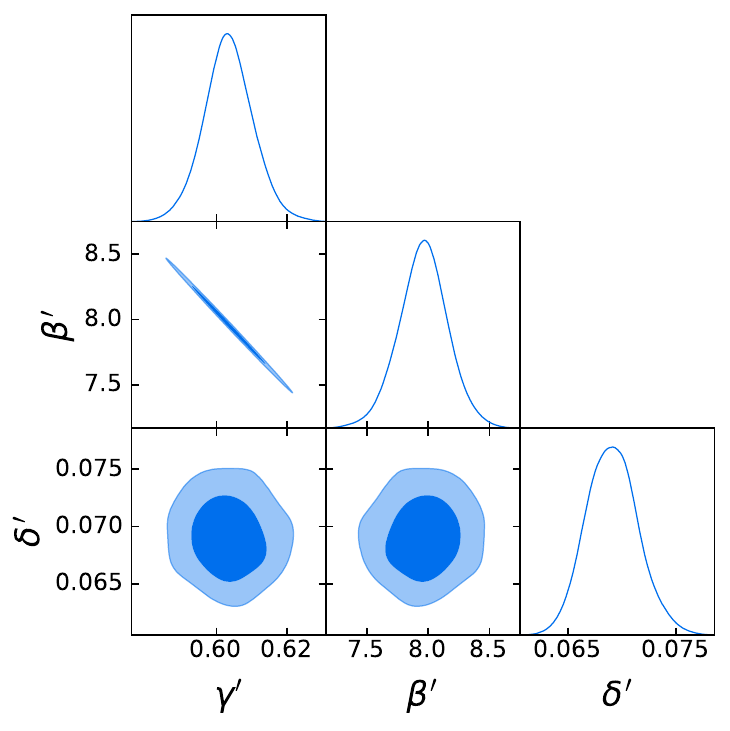}
\caption{{Corner} 
plot of the $L'_\mathrm{X} - L'_{\mathrm{UV}}$ relation, corresponding to the $F_\mathrm{X} - F_{\mathrm{UV}}$ relation, once corrected for the effects of selection and the evolution in redshift for the sample of 1065 QSOs. The resulting best-fit parameters are: $\gamma{'}=0.60\pm0.01$, $\beta{'}= 7.96  \pm 0.20 $, and $\delta{'} = 0.069 \pm 0.002$. {Here, the fiducial cosmology assumed is a flat $\Lambda$CDM model with $\Omega_M = 0.3$ and $H_0 = 70  \, \mathrm{km} \, \mathrm{s}^{-1} \, \mathrm{Mpc}^{-1}$. The brighter color indicates the probability of the occurrence of the parameters at 95\% and the darker color at 68\%.  } }
\label{fig:RL_noadd}
\end{figure}

{The} 
results of these analyses are shown in the right panels of Figures~\ref{fig:fxfuv_noadd} and~\ref{fig:fxfuv} and Figures~\ref{fig:RL_noadd} and~\ref{fig:RL}. We also note that in the right panels of Figures~\ref{fig:fxfuv_noadd} and~\ref{fig:fxfuv} the RL {correlation in the luminosity--luminosity space} is presented by assuming a certain cosmological model and corrected for selection biases and redshift evolution. Therefore, these plots are only for the purpose of showing the reached tighten relation, and they are presented as an example assuming a flat $\Lambda$CDM model with $\Omega_M=0.3$ and $H_0= 70 \, \mathrm{km} \, \mathrm{s}^{-1} \, \mathrm{Mpc}^{-1}$. 
The code and a comprehensive technical description of this method can be accessed from the Wolfram Mathematica Notebook Archive~\cite{EP_notebook}.
Before performing any cosmological analysis with our selected samples, we have also demonstrated through simulations that we are able to retrieve any assumed input cosmology for a mock QSO sample with a similar redshift and flux distribution
to our final sample. Indeed, we have generated 1065 and 1132 mock data, respectively, for the two final samples, with distributions of redshift (see Figure \ref{fig:hist_z}), fluxes, and uncertainties on fluxes drawn from the corresponding best-fit distributions of our final samples of observed data composed of 1065 and 1132 sources. Also, we have assumed a priori a cosmological model to compute the luminosities, and we have also fitted the cosmological parameters of the investigated cosmological model. Specifically, we have investigated different assumptions for the cosmological model: flat $\Lambda$CDM models with $\Omega_M = 0.3$ and $H_0=70  \, \mathrm{km} \, \mathrm{s}^{-1} \, \mathrm{Mpc}^{-1}$, with $\Omega_M = 0.1$ and $H_0=80  \, \mathrm{km} \, \mathrm{s}^{-1} \, \mathrm{Mpc}^{-1}$, with $\Omega_M = 0.5$ and $H_0=65  \, \mathrm{km} \, \mathrm{s}^{-1} \, \mathrm{Mpc}^{-1}$, and with $\Omega_M = 0.8$ and $H_0=60  \, \mathrm{km} \, \mathrm{s}^{-1} \, \mathrm{Mpc}^{-1}$. In all these cases, fixing the value of $H_0$ to the assumed one and applying the redshift correction with ``varying evolution'', we have recovered the assumed value of $\Omega_M$ within 1$\sigma$.
We here clarify that we use the notation of ``outliers'' not in a strict statistical sense but rather to refer to the QSOs that show more discrepancy from the RL relation line.
With the above-described division of the sample into bins (see also Table~\ref{tab:bins}), we can prove that there is not a particular trend or significant behavior of the slope $a$ as a function of the redshift. We here show that the trend of the slope values corresponding to the average redshift fluctuates around an average value of $a=0.60$ (see Figure~\ref{fig:binvsslope}). Some fluctuations are visible, but it is expected as the sample size is not equally divided according to the number of sources.

\subsubsection{The Parameter $\epsilon$ in the Huber Procedure}
\label{sec:epsilon}

{We can also take advantage of the parameter $\epsilon$, which is a free parameter in the Huber regression method in the range $[1, \inf)$ which can be arbitrarily fixed. This parameter controls the number of samples that should be classified as outliers: the smaller $\epsilon$ is, the more robust the Huber regression is to define outliers. We have indeed performed our selection by trying different values for $\epsilon$, and we have obtained compatible results in all the attempts. Thus, we have identified the value of $\epsilon=1.2$ as the one that leads to the cosmological result with the smallest uncertainty on $\Omega_M$ and yet a considerable statistical sample. 
The impact of the choice of $\epsilon$ on our analysis is also shown in Sections~\ref{sec:widerbins}--\ref{sec:nn}.
}

\subsubsection{The Impact of the Binning on the Data Analysis: Bins Wider in Redshift}
\label{sec:widerbins}
As anticipated, to further investigate the impact of our choice for the division in redshift bins (i.e., $\mathrm{log_{10}}(1/(1+\Delta z))$ with $\Delta z = 0.042$) (see, e.g.,~\citep{2022arXiv221102129C} for a discussion on the binning), we have also selected the QSO sample by using three other choices for the division in bins, different from the one detailed above, which are described here and in Sections~\ref{sec:optimizedbins} and~\ref{sec:nn}. 

\begin{figure}[H]
\includegraphics[width=12cm]{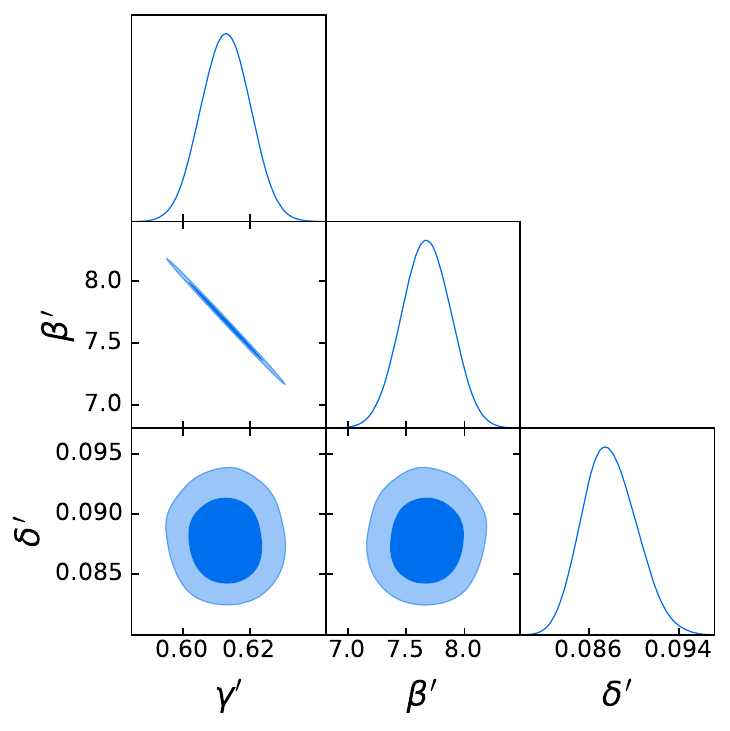}
\caption{Corner plot of the $L{'}_\mathrm{X} - L{'}_{\mathrm{UV}}$ relation, corresponding to the $F_\mathrm{X} - F_{\mathrm{UV}}$ relation, once corrected for the effects of selection and the evolution in redshift for the sample of 1132 QSOs. The resulting best-fit parameters are: $\gamma{'}=0.61 \pm 0.01$, $\beta{'}= 7.67  \pm 0.21 $, and $\delta{'} = 0.088 \pm 0.002$. {Here, the fiducial cosmology assumed is a flat $\Lambda$CDM model with $\Omega_M = 0.3$ and $H_0 = 70  \, \mathrm{km} \, \mathrm{s}^{-1} \, \mathrm{Mpc}^{-1}$. The dark region shows the 68\% of probability of the parameters at play, while the lighter blue region the 95\%.}}
\label{fig:RL}
\end{figure}

We here start from the method of dividing the sources in bins whose width depends on the redshift (see the blue box in Figure~\ref{fig:algorithm}). Specifically, we have divided the original sample according to the following prescription: $\Delta \mathrm{log_{10}}z=0.03$ if $z\leq1$, $\Delta \mathrm{log_{10}}z=0.04$ if $1 < z\leq 2$, $\Delta \mathrm{log_{10}}z=0.05$ if $2<z\leq 4$, and $\Delta \mathrm{log_{10}}z=0.06$ if $z>4$. The choice of increasing the width of the bins for higher redshifts is justified by the trend of the luminosity distance $D_L(z)$ 
Indeed, the trend of $D_L(z)$ is much steeper at low $z$ compared to the one at higher $z$, where the function $D_L(z)$ flattens; thus, we need to impose narrower $\Delta \mathrm{log_{10}}z$ at low redshifts to verify the condition that the difference $\Delta \mathrm{log_{10}}D_L$ is smaller than the intrinsic dispersion in the corresponding bin. The specific values of $\Delta \mathrm{log_{10}}z$ in each redshift range have been chosen to guarantee the fulfillment of all the required criteria detailed above: a minimum number of 10 sources that fulfill the requirement conditions explained above, a range of values of $\mathrm{log_{10}}z$ within the chosen redshift interval so that $\Delta \mathrm{log_{10}}D_L$ is smaller than
the intrinsic dispersion of the relation in the same bin, and the Anderson--Darling two-sample test that is passed at a statistical level of at least $5\%$ to warrant that the selected sample is drawn from the original one in the investigated bin. By applying this division into bins, we generate 29 bins with at least 10~sources.
{As already discussed in the previous section, also in this case the value $\epsilon = 1.2$ for the Huber regressor proved to be the one, among the several values of $\epsilon$ tested, that leads to the cosmological results with smaller uncertainties on $\Omega_M$. Furthermore, the results obtained by applying different values of $\epsilon$ are completely compatible with each other.}
Ultimately, with this procedure, we identify two final samples, as described in Point (4) of Section~\ref{sec:selection}: one comprising 1084 QSOs, which does not contain the sources in the bins in which some of the conditions are not fulfilled, and one comprising 1125, which also includes these sources (see Table~\ref{tab:bincases}).
This analysis, although a good alternative to the previous method, does not allow us to reach the golden sample {if we consider the precision on $\Omega_M$ obtained for both samples, with and without untouched sources}.

\begin{table}[H]
\caption{{Comparison} 
of results obtained from different selection approaches. The first column details the method applied to divide the initial QSO sample in bins, the second column specifies if the sample includes or not the sources in bins that do not fulfill the criteria required by our trimming analysis (see Sections~\ref{sec:selection} and~\ref{sec:widerbins}--\ref{sec:nn} for further details). The third and fourth columns report, respectively, the number of sources (N) in the considered sample after the removal of outliers and the estimated $\Omega_M$ with its 1$\sigma$ uncertainty.}
\begin{adjustwidth}{-\extralength}{0cm}
\label{tab}
\begin{tabularx}{\linewidth}{m{6cm}<{\centering}m{2.3cm}<{\centering}m{1cm}<{\centering}CCC} 
\toprule
\textbf{Method} & \textbf{Sample} & \textbf{N} & $\boldsymbol{\Omega_M \pm \Delta \Omega_M}$ & {$\boldmath{\rm z\text{-}score_{gold}}$ } 
& {$\boldmath{\rm z\text{-}score_{SNe}}$} \\ 
\midrule
Bins in $\mathrm{log_{10}}(1/(1+\Delta z))$  & {Without sources} & 1065 & $0.231 \pm 0.122$ & $0.015$ & $-0.835$\\
\midrule
& \makecell{With sources} & 1132 & $0.256 \pm 0.089$ & $0.250$ & $-0.859$\\
\midrule
Bins wider in $z$ & {Without sources} & 1084 & $0.317 \pm 0.162$ & $0.508$ & $-0.104$\\
\midrule
& \makecell{With sources} & 1125 & $0.229 \pm 0.061$ & $0$ & $-1.651$\\
\midrule
{Bins optimized in width from highest $z$} & {Without sources} & 1843 & $0.362 \pm 0.162$ & $0.768$ & $0.172$\\
\midrule
& \makecell{With sources} & 1858 & $0.285 \pm 0.110$ & $0.445$ & $-0.440$\\
\midrule
{Bins optimized in width from lowest $z$} & {Without sources} & 1965 & $0.482 \pm 0.222$ & $1.099$ & $0.664$\\
\midrule
& \makecell{With sources} & 1980 & $0.349 \pm 0.149$ & $0.745$ & $0.100$\\
\midrule
Bins centered on each QSO & {Without sources} & 811 & $0.490 \pm 0.226 $~* & $1.115$ & $0.688$\\
\midrule
& \makecell{With sources} & 825 & $0.315 \pm 0.124 $ & $0.622$ & $-0.152$\\
\bottomrule
\end{tabularx}
\label{tab:bincases}
\end{adjustwidth}
\noindent{\footnotesize{The symbol ``*'' identifies the cases in which $\Omega_M$ is not constrained. The parameter {$\mathrm{z\text{-}score_{gold}}$} is a comparison between each result and the value with the smallest uncertainty obtained in this work, which is $\Omega_{M}=0.229\pm 0.061$, while {$\mathrm{z\text{-}score_{SNe}}$} computes the compatibility of each result with $\Omega_M=0.338 \pm 0.018$ {from} 
~\citep{2022ApJ...938..110B} (see Appendix~\ref{sec:comparison} for details).}}
\end{table}
\vspace{-15pt}

\begin{figure}[H]
\begin{adjustwidth}{-\extralength}{0cm}
\centering
\includegraphics[width=.45\linewidth]{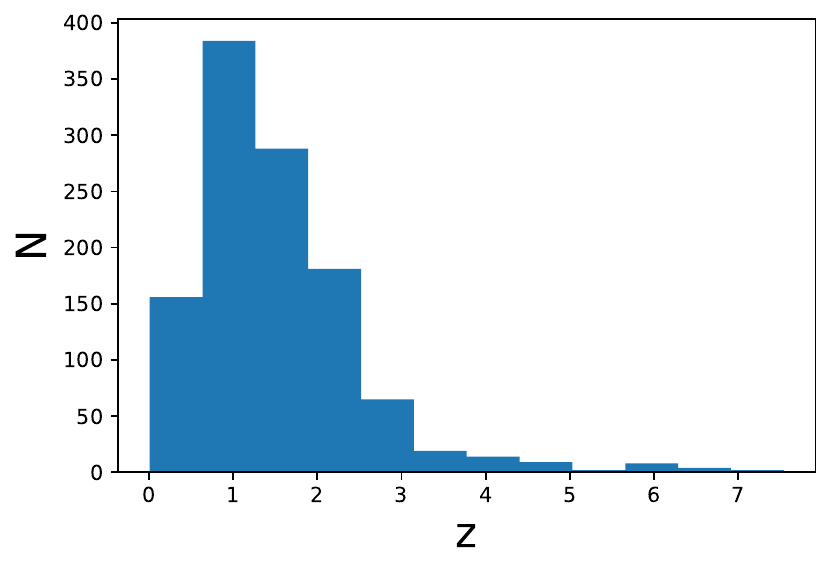} \includegraphics[width=.45\linewidth]{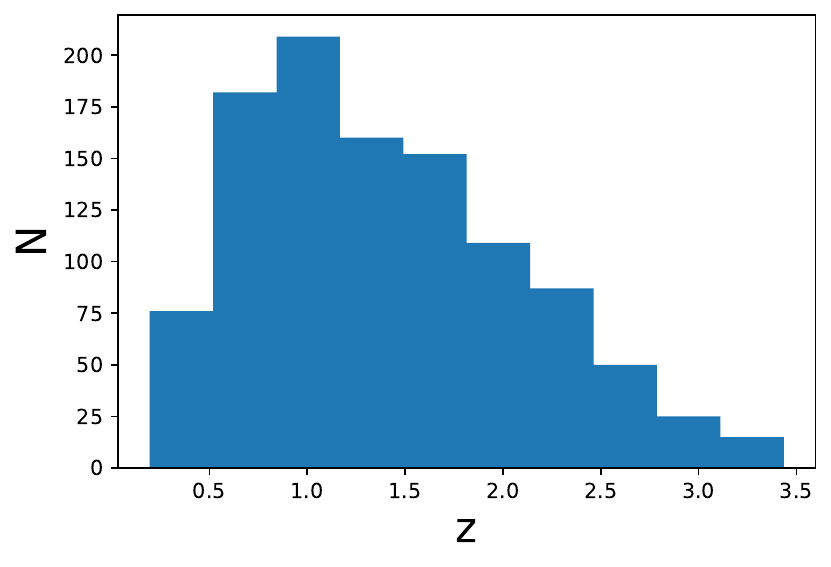}
\end{adjustwidth}
\caption{{\textbf{Left panel:} {The} 
redshift distribution of the sample with 1132 sources. \textbf{Right panel:} Same as the left panel but for the smaller sample of 1065 QSOs.} }
\label{fig:hist_z}
\end{figure}
\subsubsection{The Impact of the Binning on the Data Analysis: Optimization of the Width of Bins}
\label{sec:optimizedbins}

To avoid the arbitrary choice of the division into bins that is intrinsically needed by the two procedures outlined above, we have also developed an approach to optimize the binning. This method is based on the maximization of the number of sources (see the orange box in Figure~\ref{fig:algorithm}). More specifically, we start from the source at the highest redshift in the initial sample or the source at the lowest redshift (we have investigated both cases). We define the edge of the bins so that the maximum number of sources is included in each bin and still the same criteria of the previous procedure are fulfilled. The maximum difference in $D_L(z)$ in the bin is smaller than the intrinsic dispersion in the same bin. Also, the selected sample is still drawn from the original one in the bin according to the Anderson--Darling two-sample test with a threshold for the \emph{p}-value of 5\%.
With this procedure, we have found that the bin with fewer sources is the first, with 12~QSOs, while the most populated bin has 567 sources.
After the first bin has been identified, the same procedure is repeated to create another bin adjacent to it in redshift, and this algorithm continues until the size of the initial sample is reached. Following this recursive approach, this algorithm automatically creates bins to divide the QSO sample without any arbitrary~choice. 

When we arrive at the point where almost the whole sample is divided into bins, we could face an issue that the number of data points that still do not belong to any bin would be smaller than 10 (the assumed minimum size). Thus, it is impossible for those points to represent a reliable bin. We store those sources in a separate set called ``untouched''.
Alternatively, this could also be the case in which the number of untouched sources is larger than 10, but the condition between the $\Delta log_{10}(D_L(z))$ and the intrinsic scatter cannot be fulfilled for the set of these sources. These scenarios can occur whether we start from the source at the highest redshift or from the source at the lowest redshift. Thus, in these cases, we treat these sources that remain out of the binning division just like the untouched sources in bins without sufficient statistics. As already discussed in Section~\ref{sec:selection}, we distinguish a sample in which these QSOs are not included and another one that instead includes them. With this approach, we obtain 9 bins and 15 untouched QSOs starting from the highest redshift and 9 bins and 2 untouched sources when we start our binning procedure from the lowest redshift. 
We here notice that the number of bins is significantly smaller compared to the ${\sim}$30 bins used in the two above-described methods. This can be ascribed to the fact that in this case, we are not a priori imposing a specific division into bins. Still, the binning is automatically generated while requiring the fulfillment of the necessary conditions for our analysis. This causes a smaller number of bins, which are more populated in the range of intermediate redshifts between $z{\sim} 0.4$ and $z {\sim} 3$. This innovative procedure of optimization of bins will be described with more details and in-depth analyses in~\citep{Dainotti2023inprep}.
{As in the previous approaches, we have also tested different values for the $\epsilon$ parameter of the Huber regressor. Since the results proved to be completely compatible, we have chosen the values $\epsilon = 1.5$ and $\epsilon = 1.6$, respectively, for the methods starting from the highest redshift and the lowest redshift, as these choices guarantee the cosmological results with the smallest uncertainty on $\Omega_M$.} Using this approach, we have obtained the following final QSO samples (see Table~\ref{tab:bincases}): 1843 and 1858 sources, respectively, not including and including the sources that do not belong to any bin, for the method starting from the highest redshift, and 1965 and 1980 QSOs, for the corresponding cases when starting from the lowest redshift.
{This method allows us to reach a sample of low and high redshift, which is not covered by the other method. This allows a smaller number of untouched sources.}
Similarly to the previous method, it does not lead to the golden sample {since it does not allow us to reach the same precision as the binning division in $\mathrm{log_{10}}(1/(1+\Delta z))$}.
\subsubsection{The Impact of the Binning on the Data Analysis: A Bin Centered on Each Source}
\label{sec:nn}

Furthermore, we have also developed an additional method that allows us to completely free our analysis from the possible issues of the binning approach.
As a matter of fact, in this procedure, we do not actually associate each QSO with a specific bin. Indeed, we consider each of the initial 2421 QSO sources as the center of an interval (see the purple box in Figure~\ref{fig:algorithm}). This interval is symmetric in redshift; thus, it consists of the five sources next to the central one at lower redshifts and the five sources next to the central one at higher redshifts. In this way, each interval consists of a total of 11 sources, the minimum number we require to statistically perform a reliable fit, since we have a central source and 5 on the right-end side and 5 on the left-end side of the interval. Thus, this number must be odd by construction. Then, within each of these intervals generated around each of the initial QSOs, Steps (2), (3), and (4) outlined in Section~\ref{sec:selection} are performed. Specifically, in each of these intervals, the condition on $D_L(z)$ and the intrinsic scatter is checked, then the Huber regressor is applied, and the Anderson--Darling two-sample test is performed to verify that the selected sample is still drawn from the initial one and the same interval.

We define a source as an inlier only if the Huber regressor selects it as an inlier in all the intervals it belongs to.
As in the other approaches detailed above, we still account for the QSOs in intervals that do not satisfy the conditions of our analysis by distinguishing the two final samples in which we do not include and include these sources. 
{we have again checked that choosing different values of the parameter $\epsilon$ for the Huber regression does not impact our analysis, leading to compatible results. Thus, we have chosen $\epsilon = 1.3$ as the value that leads to the cosmological result with the smallest uncertainty on $\Omega_M$.} Hence, we obtain the following samples, summarized in Table~\ref{tab:bincases}: 811 and 825 sources, respectively, when we do not include and add the sources belonging to intervals that do not fulfill our criteria.
{Once again, this method does not lead to the golden sample if we consider the precision reached on $\Omega_M$.}


In the following, we focus on the cosmological analysis performed with the final samples obtained from the first methodology detailed, which is the one with the binning in $\mathrm{log_{10}}(1/(1+\Delta z))$ with $\Delta z = 0.042$. This is indeed the approach that, among all the methodologies investigated, leads to the cosmological result with the smallest uncertainty on $\Omega_M$. Nevertheless, we also discuss and compare the results obtained by applying the other selection procedures in {Appendix}~\ref{sec:comparison}.

\subsection{Treatment of Redshift Evolution and Selection Biases}
\label{epmethod}

Since QSOs are observed up to high redshifts, we need to correct their luminosities for selection biases and evolutionary effects~\citep{terzianevolution}, which could, in principle, distort or induce a correlation between luminosities, thus inducing an incorrect determination of cosmological parameters~\citep{Dainotti2013a}. To apply this correction, we employ the Efron and Petrosian (EP) statistical method~\citep{1992ApJ...399..345E} already used in several works~\citep{greennew,ying2019regression} for GRBs~\citep{Dainotti2013a,dainotti2015b,Dainotti2021Galax...9...95D,Dainotti2022MNRAS.tmp.2639D,Bargiacchi2023MNRAS.521.3909B,Dainotti2023arXiv230510030D} and in the QSO realm~\citep{DainottiQSO,biasfreeQSO2022,Bargiacchi2023MNRAS.521.3909B,Dainotti2023arXiv230510030D}. We use our own package to better customize it for our own analysis (see the Mathematica {notebook} accessed in 18 May 2023) 
(\url{https://notebookarchive.org/2023-05-8b2lbrh}). In this study, we apply to the obtained QSO sub-samples the procedure outlined in the above-mentioned works. In this section, we summarize the method and outcomes.

In the EP technique, we are able to determine if there is evolution among the redshift and the variables at play. With the term ``evolution'' we refer to the trend of a given variable, in this case, the luminosity with the redshift as the variation of this variable with the redshift.
The luminosities are assumed to evolve with $z$ according to $L' = \frac{L}{(1+z)^{k}}$, where $L$ is the observed luminosity, $L'$ the corresponding corrected one without evolution, and $k$ the parameter that mimics the evolution.
{The $L'$ are the corrected luminosities, where both the intrinsic evolutionary effects and selection biases have been removed, and this is the main reason we should use these values and not the uncorrected luminosities in the final computation for cosmological use.}
{From now on, with the symbol ${'}$, we indicate the new de-evolved quantities after the correction for the evolution, not only for the luminosities but also for the parameters of the RL relation (see Equation \ref{RL}).} Nonetheless, the choice of the functional form as a power-law does not affect the results~\citep{2011ApJ...743..104S,Dainotti2021ApJ...914L..40D, DainottiQSO} and hence we could also parameterize the dependence on the redshift through more complex functions. 
Then, Kendall's $\tau$ statistic is applied to identify the $k$ value that eliminates the evolution with the redshift.
In this procedure, $\tau$ is defined as
\begin{equation}
\label{tau}
\tau =\frac{\sum_{i}{(\mathcal{R}_i-\mathcal{E}_i)}}{\sqrt{\sum_i{\mathcal{V}_i}}}.
\end{equation}
where $\mathcal{R}_i$ is the rank defined as the number of points in the associated set of the $i$-source; the associated set consists of all $j$-points for which $z_j \leq z_i$ and $L_{z_j} \geq  L_{min,i}$, with $L_{min,i}$ being the minimum observable luminosity ($L_{min,i}$) at that redshift.
The EP method defines subsamples of the data, which are called ``associated sets'' and defined as containing objects, denoted with j, which should have luminosity larger than the minimum luminosity pertinent to that object and still observable according to the satellite threshold limit and the redshift should always be smaller than the redshift of a given object.
In Equation~\eqref{tau}, $\mathcal{E}_i = \frac{1}{2}(i+1)$ and $\mathcal{V}_i = \frac{1}{12}(i^{2}+1)$ are the expectation value and variance, respectively, when the evolution with redshift has been removed. As a consequence, the correlation with redshift disappears when $\tau = 0$, which allows us to obtain the value of $k$ that removes the dependence. The condition $| \tau | > n$ implies that the hypothesis of no correlation is rejected at $n \sigma$ level. We provide the 1$\sigma$ uncertainty on the $k$ value by imposing $|\tau| \leq 1$. The found value of $k$ can now be used to determine $L'$ for the total sample.
$L_{min,i}$ is computed by requiring a limiting flux. The value of this flux threshold is chosen such that the retained sample is composed at least of 90\% of the total initial sources and that it resembles the overall original
distribution according to the Kolmogorov--Smirnov test~\citep{dainotti2015b,Levine2022ApJ...925...15L,DainottiQSO,Dainotti2022MNRAS.514.1828D}. 
Indeed, we here stress that the above-described procedure employed to correct the luminosities is applied to both the X-ray and UV luminosities separately. This means that we obtain two different evolutionary coefficients, which are $k_{UV}$ and $k_X$ for $L_{UV}$ and $L_X$, respectively. The results of the application of the EP method to our two selected QSO samples, which will be described in this work, are provided in Appendix~\ref{The varying evolution}.
{Indeed, {Ref.}~\citep{DainottiQSO} has already proved that the initial QSO sample of~\citep{2020A&A...642A.150L} suffers from redshift evolution (see Figure 2 of~\citep{DainottiQSO}) and thus the luminosities need to be corrected through the EP method.}
In~\citep{2013ApJ...764...43S}, where the same evolutionary form of $(1+z)^k$ is used, they found $k_{opt}=3.0 \pm 0.5$ and corrected the luminosity function. Thus, the new luminosity function can be representative of the observed luminosity function, but it will be constructed with the local luminosities (de-evolved luminosities), and thus, they will be rescaled by the $g(z)$ functions. 

Nevertheless, from the description of the EP method, is clear that $k$ is obtained assuming a specific cosmological model, needed to compute the luminosities from the fluxes. Usually, the assumed model is a flat $\Lambda$CDM model with $\Omega_M =0.3$ and $H_0 = 70  \, \mathrm{km} \, \mathrm{s}^{-1} \, \mathrm{Mpc}^{-1}$. 
This induces the so-called ``circularity problem''.
This problem has been completely overcome for the first time by \cite{Dainotti2022MNRAS.tmp.2639D} for GRBs and \cite{DainottiQSO, biasfreeQSO2022} for QSOs, which have analyzed the trend of $k$ as a function of the cosmological model assumed a priori. More precisely, in these studies, $k$ is determined not by fixing the cosmological parameters of the assumed model, but over a grid of values of the cosmological parameters (i.e., $\Omega_M$, $H_0$, and also other parameters for models different from the flat $\Lambda$CDM one), leading to the determination of the functions $k(\Omega_M)$ and $k(H_0)$. Due to the invariance of $\tau$ under linear transformations of data, $k$ does not depend on $H_0$. However, it shows a dependence on $\Omega_M$, and thus $k(\Omega_M)$ can be applied in the cosmological fits while leaving $k$ free to vary along with the free cosmological parameters. Hence, the RL relation $\mathrm{log_{10}} L'_X = \gamma \, \mathrm{log_{10}} L'_{UV} + \beta$ can be written in terms of the evolutionary coefficients as $\mathrm{log_{10}} L_X = \gamma \, \mathrm{log_{10}} L_{UV} + \beta + k_X(\Omega_{M}) \,  \mathrm{log_{10}} (1+z) - \gamma \, k_{UV}(\Omega_{M}) \, \mathrm{log_{10}} (1+z)$.
This overcomes the circularity problem since we do not fix any cosmology a priori.
In all our computations, we employ this method, which we refer to as``varying evolution'', since it allows us to avoid the assumption of a specific value of $\Omega_M$ to correct the luminosities for this effect, for details, see the Appendix~\ref{The varying evolution}. 
We here note that the ``varying evolution'' methodology allows us to elude any degeneracy between the evolutionary coefficients, $k_{UV}$ and $k_X$, and the other fitted parameters, which are $\gamma$, $\beta$, and the cosmological parameters involved. On the other hand, letting $k_{UV}$ and $k_X$ to vary together inside a MCMC fitting, without knowing the dependence of cosmological parameters with these other parameters, would introduce degeneration. This is the reason why we determine $k_{UV}(\Omega_M)$ and $k_X(\Omega_M)$ as a step zero by applying the EP method in a completely cosmology-independent way, and then we use these functions in the cosmological fits. This issue was also already dealt with in \cite{Bargiacchi2023MNRAS.521.3909B}.
Since there is degeneracy among the $k_{UV}(\Omega_M)$ and $k_X(\Omega_M)$, we are able to determine precisely one variable if we know the other; thus, we prefer to leave the varying evolution approach with the two functions determined before we perform the cosmological fitting.
{We here also notice that the ``varying evolution'' method can be generalized to cosmological model other than the flat $\Lambda$CDM one. In this regard,~\citep{biasfreeQSO2022} show how the evolutionary coefficients in UV and X-ray behave as a function of $\Omega_M$ and $w$, the equation of state parameter, in a flat $w$CDM model and as a function of $\Omega_M$ and $\Omega_k$, the curvature density parameter, in a non-flat $\Lambda$CDM model (see their Figures 3 and 4). Following the prescription of~\citep{biasfreeQSO2022}, one can fit any cosmological model, also more complex than the standard flat $\Lambda$CDM model, by applying the ``varying evolution'' approach. Thus, it will allow us to avoid the circularity problem.}

\subsection{Cosmological Fit}
\label{cosmologicalfit}

\begin{figure}[H]
\includegraphics[width=9cm]{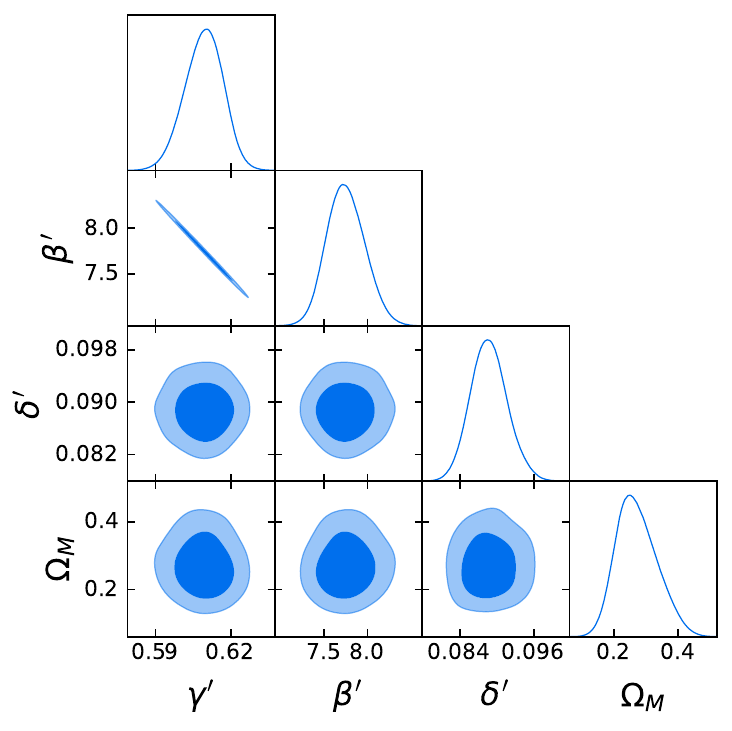}
\caption{Results obtained from the gold sample of 1132 QSOs from the cosmological fit of the flat $\Lambda$CDM model with $\gamma'$, $\beta'$, and $\delta'$ of the RL relation, corrected for redshift evolution in the luminosities and $\Omega_M$ left as free parameter together with the ones of the relation. $H_0$ is fixed to $70  \, \mathrm{km} \, \mathrm{s}^{-1} \, \mathrm{Mpc}^{-1}$  with best-fit values with 1$\sigma$ uncertainties: $\gamma{'}=0.61 \pm 0.01$, $\beta{'}= 7.8  \pm 0.2 $, $\delta{'} = 0.084 \pm 0.003$, and $\Omega_M = 0.256 \pm 0.089$. The dark region shows the 68\% of probability of the parameters at play, while the lighter blue region the 95\%.}
\label{fig:cosmo}
\end{figure}

\begin{figure}[H]
\includegraphics[width=9 cm]{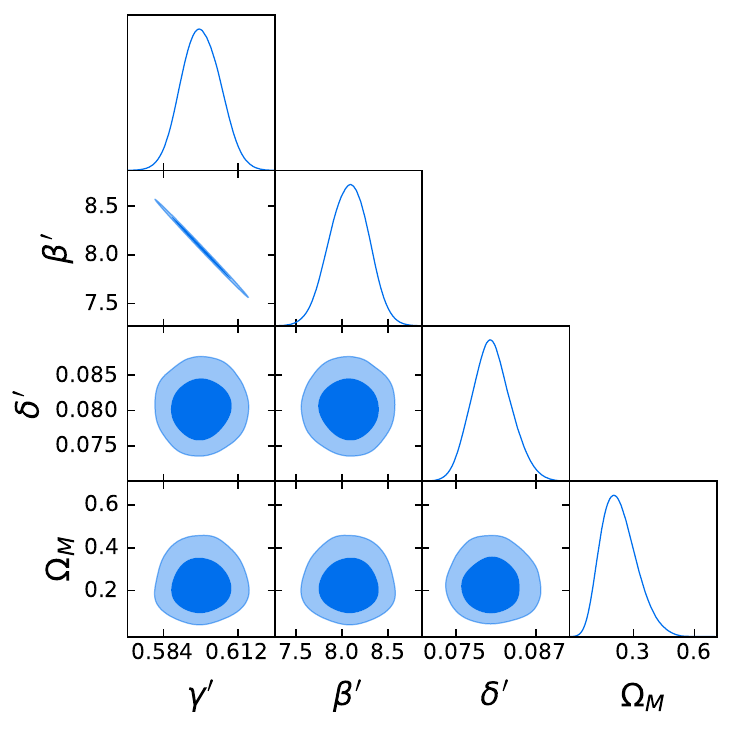}
\caption{Results from the cosmological fit of the flat $\Lambda$CDM model with $\gamma'$, $\beta'$, and $\delta'$ of the RL relation, corrected for redshift evolution in the luminosities and $\Omega_M$ left as free parameter together with the parameters of the relation for the sample of 1065 QSOs. $H_0$ is fixed to $70  \, \mathrm{km} \, \mathrm{s}^{-1} \, \mathrm{Mpc}^{-1}$  with best-fit values with 1$\sigma$ uncertainties: $\gamma{'}=0.60 \pm 0.01$, $\beta{'}= 7.9 \pm 0.2 $, $\delta{'} = 0.077 \pm 0.002$, and $\Omega_M = 0.231 \pm 0.122$. The dark region shows the 68\% of probability of the parameters at play, while the lighter blue region the 95\%.}
\label{fig:cosmo_noadd}
\end{figure}

We have employed the final QSO samples to fit with the Kelly method a flat $\Lambda$CDM model, in which we fix $H_0 = 70  \, \mathrm{km} \, \mathrm{s}^{-1} \, \mathrm{Mpc}^{-1}$, and we consider $\Omega_M$ as a free parameter with a wide uniform prior between 0 and 1 (Figures~\ref{fig:cosmo} and~\ref{fig:cosmo_noadd}). Under these assumptions, the formula for the luminosity distance $D_L$ reads (in units of Megaparsec) as 
\begin{equation}
\label{Dl}
D_L = (1+z) \frac{c}{H_{0}} \, \bigintsss_{0}^{z} \frac{d z'}{\sqrt{\Omega_{M} (1+z')^{3} + (1- \Omega_{M})}} .
\end{equation}

{{Recalling} Equation~\eqref{RLflux}, we can notice that $D_L$ can be obtained from the observed quantities $F_{UV}$ and $F_X$ as a function of the parameters of the relation. Thus, combining \mbox{Equations \eqref{RLflux} and \eqref{Dl}}, we are able to fit the cosmological free parameter $\Omega_M$ and the free parameter of the RL relation.}
Hence, we have also left $\gamma'$, $\beta'$, and $\delta'$ free to vary, we have imposed on them the uniform priors $0<\gamma'\leq 1$, $0 < \beta'< 20$, and $0 < \delta' <1$, and we have applied the best-fit cosmological likelihoods: a Gaussian likelihood for the sample of 1065~sources and a logistic likelihood for the sample of 1132 QSOs. 
Indeed, {Refs.}~\citep{snelikelihood2024,Bargiacchi2023MNRAS.521.3909B,Dainotti2023arXiv230510030D} has proven that, as the Gaussian assumption is not satisfied, the commonly used Gaussian likelihood is not the appropriate likelihood to be applied for cosmological applications of SNe Ia of \textit{{Pantheon} 
} and \textit{{Pantheon}} + samples, the whole QSO sample of~\citep{2020A&A...642A.150L}, and BAO, and that adopting the correct likelihood is crucial to reduce the uncertainties on cosmological parameters. Thus, following these works, we have checked the normality assumption for our final QSO samples uncovering that it is verified by the sample of 1065 QSOs, while it is not fulfilled by the sample of 1132 sources. Indeed, for this sample, the best-fit likelihood is a logistic one, as for the initial 2421 QSOs~\citep{Bargiacchi2023MNRAS.521.3909B}. {We here point out that the fact that the best-fit likelihood for the sample of 1065 sources is Gaussian, differently from the one of the original sample of 2421 QSOs, does not contradict the fact that the initial sample is the parent population of the selected one, as verified through the Anderson--Darling two-sample test (see Section~\ref{sec:selection}). Indeed, the two tests relate to different quantities: the Anderson--Darling two-sample test is applied to the distribution of fluxes, while the Anderson--Darling test for normality investigates the normalized residuals of luminosities. We refer to~\citep{Bargiacchi2023MNRAS.521.3909B} for a detailed analysis of the non-Gaussianity of the full sample of QSOs.} We notice that we always test the best-fit probability density function for the assumed cosmological model (e.g., $\Omega_M=0.3$, $H_0=70  \, \mathrm{km} \, \mathrm{s}^{-1} \, \mathrm{Mpc}^{-1}$) once luminosities are corrected for the evolution. Hence, we have applied the best-fit likelihoods for each sample to fit the flat $\Lambda$CDM model. 
We here note that, since the best-fit distribution for the initial QSO sample is a logistic one, the tails of this distribution cannot be neglected and the standard deviation of this logistic distribution is rather large. 
In this fitting procedure, we have also accounted for the effects of the evolution in redshift of QSO luminosities. Indeed, to fit a cosmological model, and thus explicitly show the dependence on $D_L$, we need to turn fluxes into luminosities according to $L_{X,UV} = 4 \, \pi \, D_L^2  \,F_{X,UV}$, where $D_L$ is provided by Equation~\eqref{Dl}. As anticipated in Section~\ref{epmethod}, we here have applied the most general method for correcting for this evolution, the ``varying evolution''~\citep{biasfreeQSO2022}, in which the correction varies as a function of~$\Omega_M$.

\vspace{-12pt}

\section{Results}
\label{results}

\subsection{The Gold Sample of 1132 QSOs}
\label{sec:goldsampleresults}

We here outline the main results obtained for the gold sample of 1132 QSOs, since, compared to the one of 1065 sources, this is the one that gives the best precision in terms of the cosmological results.
This sample presents a dispersion in fluxes of $\delta_{\mathrm{F}}=0.22$, which is 24\% less than the dispersion of the original sample ($\delta_{\mathrm{F}}=0.29$), {and $\delta' = 0.07$ vs. $\delta' = 0.09$ for the luminosity relation}. It still covers the whole redshift range from $z=0.009$ to $z=7.54$, and it is smaller in size compared to the initial sample (53\% of sources are discarded). We here stress that the dispersion $\delta_F$ cannot be derived from the values reported in Table~\ref{tab:bins}. Indeed, Table~\ref{tab:bins} provides the best-fit values of the intrinsic dispersion in each redshift bin investigated, while the intrinsic dispersion of the final selected sample must be computed by fitting together the sources in the whole redshift range. {More precisely, $\delta_{\mathrm{F}}=0.22$ is obtained by fitting the flux--flux linear relation $\mathrm{log_{10}}F_{X} = a \, \mathrm{log_{10}}F_{UV} + b$ on the whole redshift range covered by the selected sample.} It is also visible from Figure~\ref{fig:fxfuv} that the dispersion is reduced compared to the one of the full sample. We note that our fitted RL parameters are re-estimated after removing the ``outlying'' observations. Thus, we obtain a new set of residuals. This means that the new set of residuals is not a truncated version of the original residuals. 
Moreover, our final sample still presents the same features as the parent sample. Indeed, we have applied the Anderson--Darling two-sample test to check that the two distributions in fluxes in each bin are drawn from the same parent population of the initial sample in the same bin. This {allows us to statistically affirm compatibility with the null hypothesis} that we are not introducing biases or significant changes in the initial QSO sample from a physical point of view. 
In addition, the reduction in the sample size is not surprisingly small, since the 1048 Pantheon SNe Ia have been slimmed down from an original sample of 3473 events, with a reduction in size of the 70\% of the starting data set~\citep{scolnic2018}.
In this regard, we acknowledge that we start from an already selected QSO sample~\citep{2020A&A...642A.150L}, which has been determined from a much larger sample. Nevertheless, since we perform a cosmological analysis on QSOs, we do not start from all the original observed sources presented in the catalogs, but from the sample described in~\citep{2020A&A...642A.150L}, which contains only the sources that are standardizable cosmological candles and thus can be used in cosmological studies. 
As a further step, we have also proved that our final sample still follows the RL relation after correcting for selection effects and evolution in redshift of luminosities. Indeed, as shown in Figure~\ref{fig:RL}, we have obtained values of the slope and normalization consistent with those of the corresponding $L'_{X}-L'_{UV}$ relation for the original sample 
\citep{DainottiQSO}.

Also, we have fitted a flat $\Lambda$CDM model, {as detailed in Section~\ref{cosmologicalfit}, by fixing $H_0$ and applying the ``varying evolution'' method with the function $k(\Omega_M)$, in which the evolutionary coefficients of the EP method vary along with the free parameter $\Omega_M$ to avoid any circularity problem.} We have also obtained closed contours on $\Omega_M$ along with a significant reduction in the uncertainty on $\Omega_M$ compared to the one obtained with the whole QSO sample, which is 0.210~\citep{biasfreeQSO2022}.
Indeed, with the gold sample of 1132 QSOs, we have obtained $\Omega_M= 0.256 \pm 0.089$, as shown in Figure~\ref{fig:cosmo}, with a precision improved of 58\% compared to 0.210.
This precision is also slightly improved compared to the one reached with SNe Ia by~\citep{2011ApJS..192....1C}, which is 0.10, {even though in the case of SNe Ia $\Omega_M$ is not the only free parameter of the fit}. 
Since we are aware that QSOs at $z>3$ (see, e.g.,~\citep{2023A&A...677A.111T} for more details on the sources at $z {\sim} 3$)  show a different distribution in the flux--flux plane, compared to the sources at lower redshift (see, e.g.,~Figure 3 of~\citep{rl15} and the left and right panels of Figure~\ref{fig:fxfuvcolorbar}), we expect that the use of QSOs in bins at $z>3$ could reduce the precision on the fitted $\Omega_M$. We refer to~\citep{2023arXiv231100072R} for a theoretical analysis of the low- and high-redshift QSOs. In addition, the different cosmological roles of high and low-redshift QSOs can also be read in terms of the effects that should arise from the diversity of QSOs according to the Quasar Main Sequence~\citep{Marziani2010MNRAS.409.1033M,She2014Natur.513..210S,Negrete2018A&A...620A.118N,Dultzin2020FrASS...6...80M,Bon2020A&A...635A.151B}. Indeed, we expect a bias at higher redshift, where only extreme accretors are observed, while it may not be the case for closer QSOs, and that could affect the results. Hence, we have performed our analysis also considering only the bins at $z<3$. With this cut in redshift and including also the untouched sources, our selected sample is reduced from 1132 to 1062 QSOs. This sample constrains $\Omega_M=0.203 \pm 0.073$. As expected, the sample cut at $z<3$ improves the sensitivity on $\Omega_M$ by reducing the uncertainty by a factor of 18\%. Nevertheless, we here stress that, for our study, it is important to consider QSOs at all redshifts because the restriction to a particular redshift range in the analysis may bias the results, {as it will introduce further incompleteness in the sample}.

\subsection{The Comparison of the Two Final QSO Samples}
\label{sec:goldsampleresults_2}

As shown in Figure~\ref{fig:fxfuv_noadd}, it is also possible to achieve an increased reduction of the intrinsic dispersion, compared to that of the 1132 QSOs, if we discard the untouched sources (see Section~\ref{sec:selection}). Following this approach, we have defined a QSO sample with 1065 sources with reduced redshift coverage between $z=0.2$ and $z=3.4$ and intrinsic dispersion of the flux relation $\delta_{\mathrm{F}} = 0.18$, which is 18\% less than the dispersion of the 1132~sources and 38\% less than the dispersion of the original sample.
Moreover, as for the sample of 1132 QSOs, the 1065 sources still present the same features as the parent sample, as {tested through} the Anderson--Darling two-sample test, and recover the RL relation, as shown in Figure~\ref{fig:RL_noadd}.
Compared to the gold sample of 1132 sources described above, which yields $\Omega_M= 0.256 \pm 0.089$, the sample of 1065 QSOs, results in $\Omega_M= 0.231 \pm 0.122$, as presented in Figure~\ref{fig:cosmo_noadd}.

\begin{figure}[H]
\begin{adjustwidth}{-\extralength}{0cm}
\centering
\includegraphics[width=.45\linewidth]{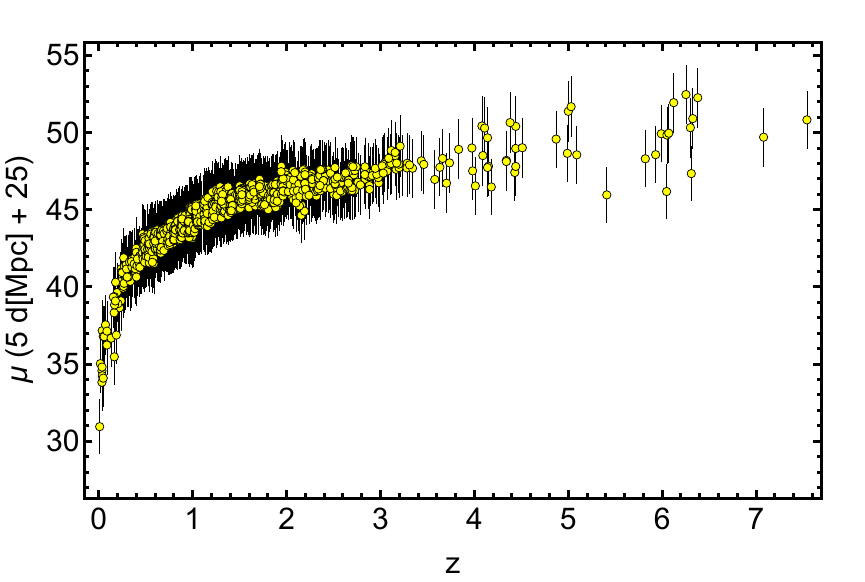} \includegraphics[width=.45\linewidth]{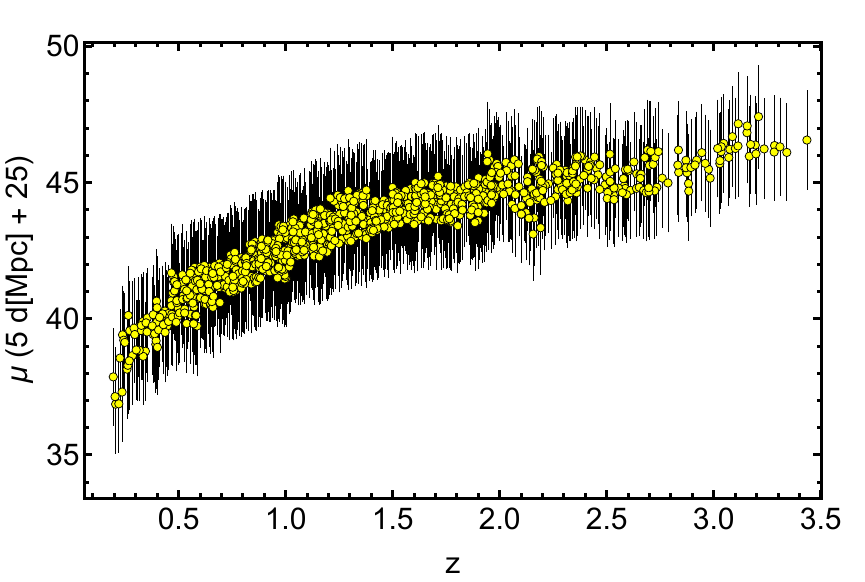}
\includegraphics[width=.45\linewidth]{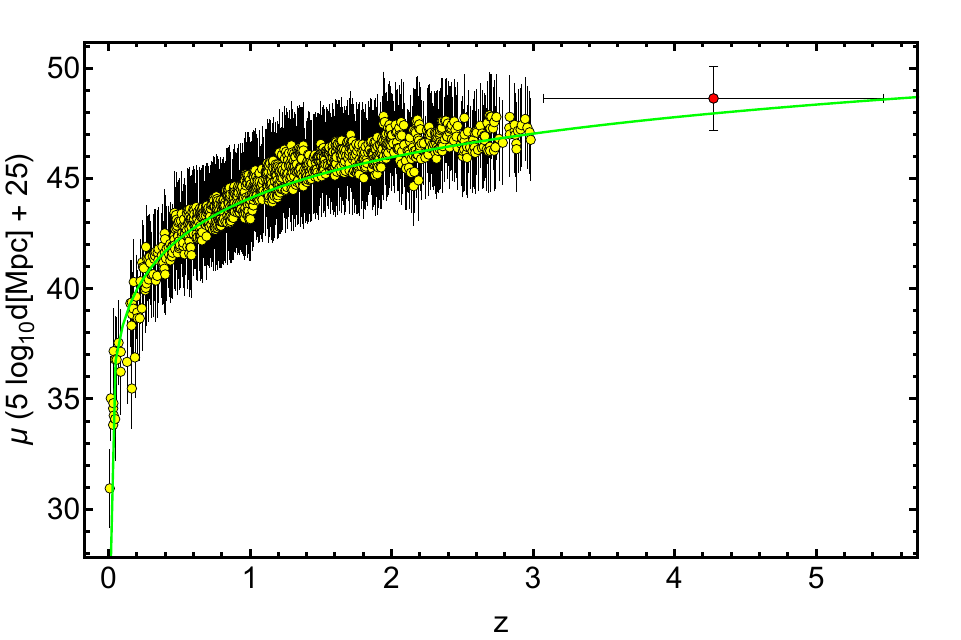} \includegraphics[width=.45\linewidth]{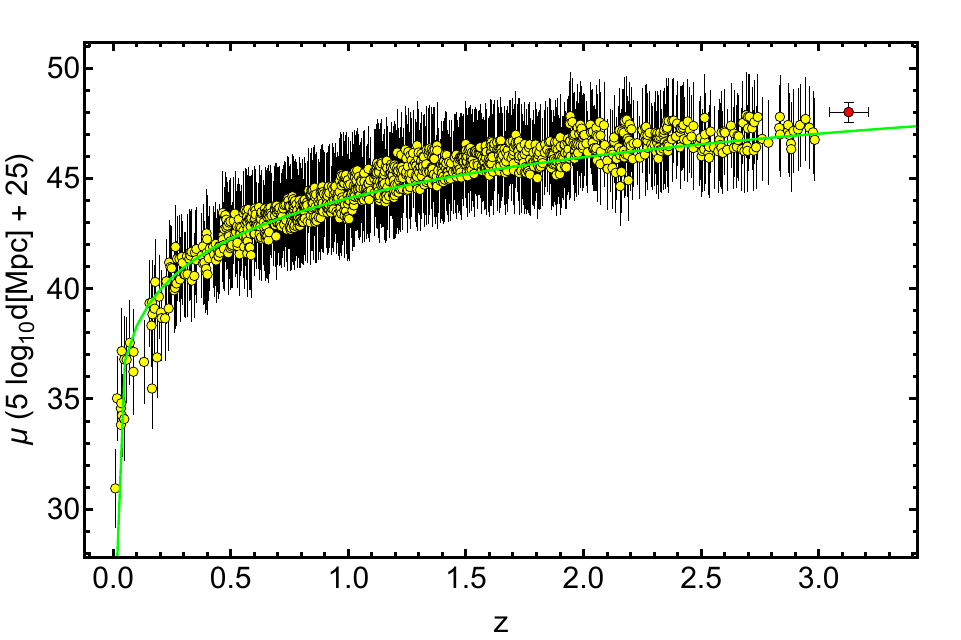}
\end{adjustwidth}
\caption{\textbf{Upper left panel:} {The} 
Hubble diagram of the gold sample of 1132 QSOs (in bright yellow) derived by assuming a flat $\Lambda$CDM model with $H_0$ fixed to $70 \, \mathrm{km} \, \mathrm{s}^{-1} \, \mathrm{Mpc}^{-1}$ and $\Omega_M= 0.256 \pm 0.089$, as obtained by fitting this sample. The error bars represent the statistical 1$\sigma$ uncertainties. \textbf{Upper right panel:} Same as the left panel but for the sample of 1065 QSOs, for which the best-fit value of $\Omega_M$ is $0.231 \pm 0.122$. \textbf{Bottom left panel:} {Same as above, but the QSOs at $z>3$ are averaged into one data point (shown in red).} \textbf{Bottom right panel:} {Same as above, but the QSOs at $3<z<3.3$ are averaged into one data point (shown in red). Plots in the bottom part mimics the analysis performed by~\citep{2022A&A...663L...7S}. The error bars on the averaged data points are computed as simple standard deviation. The green line is a plotted theoretical luminosity distance for a flat $\Lambda$CDM model with $\Omega_M=0.3$.}}
\label{fig:hd}
\end{figure}

A remarkable difference between the gold samples of 1132 and 1065 is that in the smaller sample, the highest redshift is $z=3.435$. The larger sample of 1132 contains 45~more QSOs, which are in the range of $3.435 \leq z \leq 7.5413$ and are absent in the 1065 QSO sample. These additional high-$z$ QSOs are distributed according to the redshifts as follows: 9 QSOs within $3.435<z\leq 4$, 19 within $4<z \leq 5$, 6 within $5<z \leq 6$ and 11 within $6<z \leq 7.5$. The left and right panels of Figures~\ref{fig:fxfuvcolorbar} and~\ref{fig:lxluvcolorbar} show these differences since the color bar axes are different in the two panels: the left panel has a redshift ranging from 0.009 to 7. The right panel has a redshift ranging from 0.1948 to 3.435.
This difference in the redshift coverage is also clearly visible {looking at Figure~\ref{fig:hist_z} that shows the redshift distributions of the two samples and} when comparing the top left and right panels of Figure~\ref{fig:hd}. These figures show indeed the Hubble diagrams (i.e., distance modulus vs. redshift) for both our two final samples (yellow points) along with the 1$\sigma$ uncertainty on the distance modulus. {On the bottom panels of Figure~\ref{fig:hd}, we compare if high-z data in our analysis show a significant deviation from the flat $\Lambda$CDM model with $\Omega_M=0.3$. This analysis is similar to the one presented by~\citep{2022A&A...663L...7S}, where a 4-$\sigma$ incompatibility was claimed for a set of sources in the redshift range $3<z<3.3$. Moreover, {Ref.}~\citep{2022A&A...663L...7S} obtained lower luminosity distance at high-$z$ than predicted by flat $\Lambda$CDM with $\Omega_M=0.3$. For the purpose of comparison with the aforementioned work, we averaged redshifts and the best-fit luminosity distances for the sources of our gold sample in the same redshift range. We computed the error bars as a simple standard deviation. The results are shown on the bottom-right panel of Figure~\ref{fig:hd}. Noticeably, we obtain compatibility within $<2 \sigma$, and our averaged point (red) lays above the $\Lambda$CDM with $\Omega_M=0.3$ (green line). The results presented here and in~\citep{2022A&A...663L...7S} are a reflection of obtained values of $\Omega_M$ in treatments with and without correction for evolution. The high luminosity distance at high-z in our sample leads to a slightly smaller value of $\Omega_M$ than 0.3. Similarly, when no correction for selection bias and redshift evolution is applied, one gets $\Omega_M$ going towards the value of unity since the value of luminosity distance is much smaller than the one predicted by $\Omega_M=0.3$ as obtained by \cite{2022A&A...663L...7S}. At the bottom left panel of Figure~\ref{fig:hd}, we performed the same analysis as above. The only difference is that we averaged all the sources with z > 3. We deduct the same conclusion, but here, the averaged point is compatible with the $\Lambda$CDM ($\Omega_M=0.3$) within $< 1 \sigma$.}

As already explained in Section~\ref{sec:goldsampleresults}, we have also tested to what extent the precision on $\Omega_M$ is worsened by the inclusion of QSOs at $z>3$. To this end, we have cut our sample, retaining only sources at $z<3$ and discarding the untouched sources. This procedure trims the sample from 1065 to 1040 sources. This cut sample constraints $\Omega_M= 0.180 \pm 0.082$, with an uncertainty reduced by a factor 33\%, compared to the sample of 1065 QSOs in the whole redshift range. Nonetheless, as stressed above, it is important for our analysis to consider the QSO sample at all redshifts to avoid possible biases in our results.

\vspace{-12pt}
As just stressed, we can notice that the sample composed of 1065 sources presents a smaller intrinsic dispersion compared to the one of 1132 QSOs: $\delta_\mathrm{F} = 0.18$ vs. $\delta_\mathrm{F} = 0.22$ for the flux relation, and $\delta' = 0.07$ vs. $\delta'=0.09$ for the luminosity relation. This is due to the fact that the smaller sample is obtained only by considering inliers determined through the Huber regression, while the larger sample includes sources at low ($z<0.2$) and high ($z>3.4$) redshift, for which, due to the insufficient statistics of the corresponding redshift bins, was not possible to perform the fit and thus remove outliers. The addition of these sources, which cannot be identified either as inliers or outliers, increases the dispersion {of points}, as clearly visible from the comparison of Figures~\ref{fig:fxfuv_noadd} and \ref{fig:fxfuv}. Despite this reduced intrinsic dispersion, the 1065 QSOs lead to a larger uncertainty on $\Omega_M$, 0.12, compared to the uncertainty of 0.09 reached by the sample of 1132 sources. 
The reason is that the latter sample covers the whole redshift range of the initial QSO sample, filling also the low-redshift interval that allows anchoring the Hubble diagram to the zero-point near $z=0$, thus better constraining the value of the matter density today (at $z=0$). The difference in redshift coverage of the two QSO samples is also reflected by the different best-fit cosmological likelihoods: the larger sample is better fitted with a logistic likelihood, as the original QSO sample, since it resembles the initial 2421 sources in the redshift coverage, while the smaller sample fulfills the Gaussianity assumption as it ranges over a reduced redshift interval.
Based on all these results, we here claim that with more QSOs at low and high redshift ($z<0.2$ and $z>3.4$), that would allow us to remove outliers in these redshift ranges, we could reach a reduced intrinsic dispersion and smaller uncertainty on $\Omega_M$, which marks the relevance of future QSO surveys and observations.
In Appendix~\ref{sec:comparison}, we show how our main cosmological results are completely independent of the binning~procedure.

\subsection{The Advantage of Using the Huber Regression Technique versus the Standard Fitting Methods}
\label{sec:hubergain}
As anticipated in Section~\ref{sec:selection}, we have also estimated the gain achieved with the use of the Huber regressor by comparing the results obtained with the Huber technique with the ones obtained with a traditional fitting method. To this end, we have applied the sigma-clipping procedure (see~\citep{Dainotti2023inprep} for details to select the final QSO samples), and also, we have performed the cosmological analyses of these new samples. More specifically, we have chosen the threshold of the sigma-clipping so that the size of the new final samples (with and without the addition of the untouched sources) is similar to the size of the corresponding samples obtained using the Huber algorithm. This allows us to be consistent with our analysis with the Huber approach. Specifically, by choosing a threshold for the sigma-clipping of 1.5, we have obtained a sample of 1078 QSOs when we do not include the untouched sources (compared to the 1065 of the Huber case) and a sample of 1145 QSOs when we include the untouched sources (compared in the 1132 of the Huber case).
We have also fitted the flat $\Lambda$CDM model with $\Omega_M$ free to vary, obtaining $\Omega_M =0.244 \pm 0.132$ for the sample of 1078 sources and $\Omega_M = 0.307  \pm 0.106$ for the larger sample of 1145 QSOs.
Hence, we can notice that the values of $\Omega_M$ obtained with the traditional technique and with the Huber regressor are compatible within 0.07$\sigma$  (for the sample without additional sources)  and within 0.4$\sigma$ (for the sample with additional sources), while the uncertainties on $\Omega_M$ are reduced by 8\% when we employ the Huber method (for the sample without untouched sources)  and by 16\% (for the sample with untouched sources).
Thus, this comparison shows that the best-fit of $\Omega_M$ is unchanged, as expected, but the application of the Huber selection technique allows us to improve the precision on $\Omega_M$ since it is able to better detect outliers of the RL relation and better determine the ``true'' parameters of the relation itself, thus selecting a QSO sample that is more powerful to infer cosmological~parameters.

\subsection{The Comparison of These New Gold Samples (with and without the Untouched Sources) with the RL Relation for the Total Initial Sample}
We here detail the comparison between the correlation in the de-evolved $L'_{UV}-L'_X$ plane with the RL relation valid for the total sample and not corrected for selection biases. In Figure~\ref{fig:cosmo}, the best-fit parameters obtained when leaving also $\Omega_M$ free to vary are $\gamma{'}=0.61 \pm 0.01$, $\beta{'}= 7.8  \pm 0.2 $, $\delta{'} = 0.084 \pm 0.003$, and $\Omega_M = 0.256 \pm 0.089$ for the sample of 1132 QSOs.  Without applying any correction for redshift evolution and selection biases, {Ref.}~\cite{2020A&A...642A.150L}
found the following values: $\gamma = 0.586 \pm 0.061$ and $\delta = 0.21 \pm 0.06$ with a simple forward fitting method. Hence, the slope obtained with our computation is compatible within $0.39$$\sigma$ with the one of the empirical RL relation, while the value of the intrinsic scatter is reduced by 60\%. 
Interestingly, the smaller sample of 1065 QSOs yields the following best-fit parameters: $\gamma{'}=0.60 \pm 0.01$, $\beta{'}= 7.9 \pm 0.2 $, $\delta{'} = 0.077 \pm 0.002$, and $\Omega_M = 0.231 \pm 0.122$ (see Figure~\ref{fig:cosmo_noadd}). In this case, the slope of the RL relation and that derived with our approach are compatible within 0.23$\sigma$ and the value of the intrinsic scatter is reduced by 63\%.  

Remarkably, if we consider a flat $\Lambda$CDM model in which both the value of $\Omega_M$ and $H_0$ are fixed, the smaller sample composed of 1065 QSOs, shown in Figure~\ref{fig:RL_noadd}, has similar values of the best-fit parameters: $\gamma{'}=0.60\pm0.01$, $\beta{'}= 7.96  \pm 0.20 $, and $\delta{'} = 0.069 \pm 0.002$. In this case, the slope of the RL relation and that obtained with our approach are compatible within 0.23$\sigma$ and its intrinsic scatter is reduced by 67\%. Similarly, for the first and larger sample, shown in Figure~\ref{fig:RL}, the resulting best-fit parameters are: $\gamma{'}=0.61 \pm 0.01$, $\beta{'}= 7.67  \pm 0.21 $, and $\delta{'} = 0.088 \pm 0.002$. This shows that the slope is again within $0.39$$\sigma$, similarly to the case in which instead $\Omega_M$ is free to vary. In this case, the intrinsic 
scatter is reduced by 58\%. We here note that the slope is degenerate with the normalization, but we have checked that, when we remove the degeneration by scaling the variables, the results remain compatible.
{We note that fixing $\Omega_M$ or leaving it free to vary allows a difference of 2$\sigma$ in the intrinsic scatter, where the scatter is smaller when all parameters are fixed. We expect indeed smaller values for fewer degrees of freedom.}

\subsection{The Need for This Analysis and the Interpretation of Results from a Physical Point of View}
\label{sec:physicalinterpretation}

To better understand the origin of a larger intrinsic dispersion in the QSO sample of~\citep{2020A&A...642A.150L}, we have compared the kurtosis of our sample to that in~\citep{2020A&A...642A.150L}. The kurtosis is indeed the fourth standardized moment, which identifies extreme values in the tails of the distribution compared to Gaussian tails. As reported in~\citep{Bargiacchi2023MNRAS.521.3909B}, the full QSO sample shows a kurtosis of ${\sim} 0.8$. On the other hand, for our final sample, we obtain a kurtosis of ${\sim}-$0.2 (with and without correction for evolution). Nevertheless, the larger kurtosis of the sample of~\citep{2020A&A...642A.150L} is not the only cause of the larger intrinsic scatter of this sample.
Indeed, the kurtosis in the whole data set is not very large. The larger intrinsic scatter in the whole sample means that the standard deviation (not the kurtosis) is larger. It is quite straightforward that the standard deviation (intrinsic scatter) will be smaller in our reduced sample. This is the reason why we perform and build the full methodology of all this procedure.

We here would like to stress that, since the physics and the processes that induce the X-UV relation in QSOs are not known, we are not yet able to physically explain all the observational features of the QSO sample used in cosmology. Hence, we cannot provide a physical reason that would force an upper limit of the intrinsic dispersion. Nevertheless, recently, {Refs.}~\citep{2022A&A...663L...7S,2023arXiv231208448S} has proven that several factors contribute with different percentages to the overall dispersion of the total sample and that the true intrinsic dispersion is much smaller than the one actually observed. Specifically, these factors are the following: the use of X-ray photometric measurements instead of the spectroscopic ones, the intrinsic variability of the sources, and the inclination of the torus of QSOs.
Analyzing the contribution of each of these factors, {Ref.}~\citep{2022A&A...663L...7S} has shown that, for a sample of QSOs at $z {\sim} 3$ with high-quality observations, the intrinsic dispersion is only 0.09 dex, which is completely ascribed to the intrinsic variability of QSOs and geometry effects of the sources. 
The same analysis is performed for the full QSO sample of~\citep{2020A&A...642A.150L} in~\citep{2023arXiv231208448S}. In this case, the results are as follows: the intrinsic variability produces 0.08 dex of the intrinsic dispersion, with a larger contribution at low luminosities and a smaller one at high luminosities, while the inclination contributes to 0.08 dex assuming a torus with an opening angle, measured from the disc surface (see Figure 4 of \citep{2023arXiv231208448S} for a graphical representation), of 30° (this value of 0.08 dex is lowered for larger opening angles). The use of photometry instead is negligible. We refer to~\citep{2023arXiv231208448S} for a detailed description of these contributions to the dispersion of the RL relation, but we here notice that the contribution of the inclination is computed from mock simulations, also taking into account the limb-darkening effect. Indeed, the authors start with the simple assumption of the absence of an absorbing torus. In this case, the inclination angle $\theta$ is randomly extracted from a distribution that is uniform in cos $\theta$. Then, they improve the accuracy of the model by introducing an obscurer. This way, using a  mock sample of {100000} 
QSOs, they derive a contribution of 0.08 dex to the observed dispersion from the inclination. In the end, the intrinsic dispersion proves to be 0.09 dex for the sub-sample of QSOs at $z {\sim} 3$ analyzed in~\citep{2022A&A...663L...7S} and ${\sim} 0.11$ for the whole QSO sample studied in~\citep{2023arXiv231208448S}.
The same value of 0.09 dex for the intrinsic dispersion is also obtained in~\citep{Dainotti2023inprep}, in which QSO sub-samples are selected in redshift bins through the sigma-clipping technique by retaining only the sources that better follow the RL {correlation}. Moreover, the value of ${\sim}$0.09 dex is recovered in~\citep{Dainotti2023inprep} independently of the redshift interval investigated. This shows that trimming the sample by selecting only QSOs closer to the ideal RL relation line intrinsically removes low-quality data, which are thus outliers, reducing the dispersion towards its true intrinsic value.
The value of 0.11 is much smaller than the actual observed dispersion of the full QSO sample, which is ${\sim} 0.2$ dex. This means that the current quality of the X-ray and UV flux measurements is not yet sufficient to reveal the true (very small) intrinsic dispersion of the relationship leading to an observed dispersion that is larger than the real one.
This proves that selecting the QSO sample to reduce the intrinsic dispersion, as in our procedure, allows us to build a sample much more similar and faithful to the one that properly follows the X-UV relation.

To further comment and interpret our results from the physical point of view of the RL relation, we have also investigated for each binning approach the compatibility between the best-fit values of the slope $a$ in each bin and the mean slope, averaged from the obtained slopes in all bins. To this end, we have computed the corresponding ``z-score'' parameter defined as $(a_i-<a>)/\sqrt{\Delta^2 a_i+ \Delta^2 <a>}$, where $a_i$ is the slope of the $i$-th bin and $<a>$ is the mean value of the slope calculated over all bins. This analysis shows that in all the binning cases the slopes in each bin are consistent with the mean value of the slope in $< 3 \sigma$. Indeed, the worst z-score is the one of the case with bins in $\mathrm{log_{10}}(1/(1+\Delta z))$, which has a minimum of $-$2.4 and a maximum of 2.6, while the best case is the one with the method of optimization of bins starting from the lowest redshift, for which the z-score is between $-$1.0 and 1.1. 

In addition, possible improvements on the sample could depend on the diversity of QSOs according to the Quasar Main Sequence~\citep{Marziani2010MNRAS.409.1033M,She2014Natur.513..210S,Dultzin2020FrASS...6...80M,Negrete2018A&A...620A.118N}. One could expect that at higher redshift we have a bias seeing only extreme accretors, while it may not be the case for closer QSOs, and that could affect the results. The selection of a sub-sample that includes only high accretors, may improve this bias~\citep{Negrete2018A&A...620A.118N}. {This means that it would be interesting to perform an analysis on only extreme accretors, thus selecting only high accretors also at low z.} Of course, one must be careful not to include the false candidates (see, for example,~\citep{Bon2020A&A...635A.151B}) Although this is a very interesting topic, this analysis goes beyond the scope of this work.

In summary, our aim is to show that, if a proper sample with a reduced intrinsic dispersion is defined, then QSOs can be used as standalone probes with the precision of SNe Ia. Our gold QSO samples can help to reveal physical properties common to these sources in order to identify a QSO sample driven by fundamental physics. However, the investigation of these physical properties goes beyond the scope of the current paper. Overall, we are discarding outliers, identified this way by the robust Huber regressor, to define a QSO sample that is able to constrain $\Omega_M$ with unprecedentedly high precision. The Huber algorithm is based on the identification of the sources that follow the true slope of the relation (inliers) and the ones that do not (outliers). In the end, the QSOs that are detected as outliers, even with measurement errors smaller than the intrinsic dispersion observed, {could} be affected by some observational problems as they do not follow the RL relation. An example of such an issue could be the host extinction for photometric measurements, where the lack of knowledge of the spectrum does not allow for a precise determination of the QSO's unabsorbed luminosity.

\section{Summary, Discussion, and Conclusions}
\label{conclusions}

To uncover the ultimate QSO sample that can be used as a powerful cosmological tool,
we started from the most comprehensive and up-to-date data set for cosmological studies~\citep{2020A&A...642A.150L}, comprising 2421 sources spanning from $z=0.009$ to $z=7.54$, and we ushered in an {original} approach.
We stress that, differently from other works~\citep{2020A&A...642A.150L,2022MNRAS.515.1795B}, we use the full sample at all redshifts and we also correct for redshift evolution to ensure the utmost accuracy and reliability for cosmological applications.
Our procedure to determine such a gold sample is general and versatile, and hence can also be used for other probes (e.g., GRBs) and larger samples.
Additionally, this method is completely model-independent and thus, it avoids the circularity problem.
Indeed, we apply our technique to the linear relation between the logarithms of QSO fluxes in UV and X-ray, that do not depend on the choice of a specific cosmological model, contrary to the case of luminosities. 
We employ the robust Huber regression which allows us to reduce the intrinsic scatter of the $F_X-F_{UV}$ relation by removing the sources that are identified as ``outliers''~\citep{owen2007robust}, a notation that we use not in its strict statistical sense but rather to refer to the QSOs that show more discrepancy from the best-fit relation line.
The novelty in our approach lies in harnessing the remarkable power of the Huber fitting method to unearth the optimal QSO sample.
Indeed, as detailed in Section \ref{sec:hubergain}, the employment of the Huber regressor in place of a standard fitting technique allows us to determine $\Omega_M$ in a flat $\Lambda$CDM model with increased precision. Indeed, the Huber procedure better distinguishes the outliers of the RL relation, leading to a QSO final sample that better follows this relation {and} has more constraining power on the cosmological parameters.
We here outline our main results and draw conclusions.

\begin{enumerate}
\item { {The strategy in the selection of the QSO gold sample}}
. Since our main challenge is to constrain cosmological parameters, such as $\Omega_M$, we strive not only to reach the smallest dispersion of the $F_X-F_{UV}$ relation but also to keep a statistically sufficient number of sources in each redshift bin. This guarantees that the fitting is still possible from a statistical point of view. Hence, we have found a compromise between these two factors 
which are antagonistic. 
The optimal number of sources found is 1132 since these sources fulfill all the following required criteria: the minimum number of sources (i.e., 10 in our case), the Anderson--Darling two-sample tests in each bin, and the requirement on the distance luminosity that should be negligible compared to the dispersion of the relation in flux in each bin. 
This sample of 1132 QSOs presents $\delta_{\mathrm{F}} = 0.22$ and still covers the whole redshift range of the original sample, from $z=0.009$ to $z=7.54$.
The intrinsic dispersion of the flux relation can be even reduced if we discard the sources belonging to redshift bins with not enough statistics (i.e., <10), the ``untouched'' sources, to perform the Huber regression. By applying this choice, we have defined a QSO sample with 1065 sources with reduced redshift coverage between $z=0.2$ and $z=3.4$ and intrinsic dispersion of the flux relation $\delta_{\mathrm{F}} = 0.18$.

\item { {Comparison with the original RL relation and cosmological results.}} We have proven that the RL relation is still verified by our two final samples, once accounted for the redshift evolution. We have also used the obtained QSO gold samples for cosmological application to derive $\Omega_M$ by fitting a flat $\Lambda$CDM model leaving contemporaneously free both $\Omega_M$ and the RL relation parameters, $\gamma{'}$, $\beta{'}$, and $\delta{'}$, while fixing $H_0 = 70  \, \mathrm{km} \, \mathrm{s}^{-1} \, \mathrm{Mpc}^{-1}$. We have performed this fit by using the best-fit proper likelihoods for the samples, which proved to be a logistic one for the sub-sample of 1132~QSOs and a Gaussian one for the 1065 sources. We have also performed the fit by applying the correction for the redshift evolution as a function of $\Omega_M$.
The gold sample of 1132 QSOs has provided $\Omega_M=0.256 \pm 0.089$, whereas the sample with 1065 sources has led to $\Omega_M=0.231 \pm 0.122$. Hence, we have reached a precision improvement of 58\% compared to the one obtained with the whole QSO sample (i.e., 0.210, see~\citep{biasfreeQSO2022}). 
Moreover, these values are compatible with the current value of $\Omega_M=0.338 \pm 0.018$~\citep{2022ApJ...938..110B}, and hence in agreement with the expected value of the current matter density.
Additionally, the obtained values of $\Omega_M$ are compatible in ${\sim} 1 \sigma$ with the one reported in~\citep{DainottiGoldQSO2023}, in which a non-binned analysis, independent from the one here presented, is performed to select the QSO sample and constrain $\Omega_M$.
We here point out that our analysis is not biased or induced by any circular reasoning. The sample is trimmed by reducing the uncertainties in the flux--flux relationship, which does not depend on the cosmological parameters once we bin the data (see Section~\ref{sec:selection}), but rather on the intrinsic scatter of this relation. Our results show that, after restricting our attention to the probes for which the intrinsic scatter is small, we obtain a substantially improved precision of the estimated cosmological parameter.
\item { {The impact of bin division on cosmology.} }The above-detailed results have been obtained by trimming the initial QSO sample in bins of $\mathrm{log_{10}}(1/(1+\Delta z))$, as described in Section~\ref{sec:selection}, since this is the approach that leads to the best cosmological results. Nevertheless, to further investigate the impact of this choice for the division into bins and to free our analysis from the arbitrariness and possible issues of the binning procedure, we have also performed our study by applying the three different selection methodologies outlined in Sections~\ref{sec:widerbins}--\ref{sec:nn}. We have thus proved that our results do not depend on the specific methodology employed to select the QSO sample. Indeed, as detailed in Appendix~\ref{sec:comparison} and Table~\ref{tab:bincases}, the values of $\Omega_M$ obtained in all the cases investigated are compatible within $1.2 \, \sigma$  and they are also consistent with $\Omega_M=0.338 \pm 0.018$ reported in~\citep{2022ApJ...938..110B}. This compatibility with the most recent value of $\Omega_M$ measured from SNe Ia ensures that our analyses recover the expected cosmology independently of the binning approach considered. Moreover, the comparison among the results obtained from different procedures
has also shown that larger sample sizes shift $\Omega_M$ towards values of ${\sim} 0.5$, as expected from~\citep{biasfreeQSO2022}. Additionally, this extended analysis has also suggested that the low-$z$ QSOs tend to lower the value of $\Omega_M$.  
Finally, the employment of these several independent approaches to select the final QSO samples has led to an extensive knowledge of the QSO selection and trimming procedure. Furthermore, leveraging the advantages of the methods investigated, we have avoided our analyses and results being biased or distorted by the arbitrary choice of a fixed binning. The outcomes of all these approaches have established the validity of our results and the robustness of our analysis.

\item { {The need for a larger sample and a physical interpretation.}} Based on all these considerations, we here point out that we would need a much larger sample with these properties to reach the current value of precision obtained with SNe Ia, which is $\delta_{\Omega_M}=0.018$. This situation is similar to the one occurring for GRBs~\citep{Dainotti2022MNRAS.514.1828D} where the precision of SNe Ia in~\citep{2011ApJS..192....1C} is reachable now, while for an improved precision as the one the SNe Ia have today, we would need to wait for two other decades. Nevertheless, if a common emission mechanism or properties in this sample will be driven by fundamental physics the waiting time for reaching this precision would be considerably improved. We here point out a parallel case for GRBs: if the plateau emission in the Platinum sample is driven by the magnetar, a given sample of GRBs with peculiar magnetic fields and spin period can drive the standard set. Here, the gold QSOs can help to reveal the physical meaning of these properties.
Indeed, {the current paper can allow us to identify a QSO sample, highlighted by our statistical procedure, that is driven by fundamental physics.} However, this investigation goes beyond the scope of the current paper and will be analyed in a forthcoming paper.

\end{enumerate}

In conclusion, we have shown that QSOs alone with the RL relation can now be promoted to reliable standard candles to measure cosmological parameters, such as $\Omega_M$, when a gold sample is defined.
In this framework, we are able to constrain  {the cosmological parameter $\Omega_M$} with significant precision at high redshift, up to 7.5.
This ushers a new era of QSOs as effective standard candles, in which the efforts of the QSO community can be driven to understand and delve into the differences between this gold sample and the total~one.

A scientific revolution is underway. QSOs, once enigmatic cosmic entities, now stand tall as standard candles illuminating the darkest corners of our Universe. The gold sample is our key to unlocking their true potential, propelling us towards an improved precision in cosmological studies. The journey has just begun, and the tantalizing mysteries that lie ahead beckon us to delve deeper.

\vspace{6pt}

\authorcontributions{
M.G.D. and G.B. contributed equally to this work.
Conceptualization: M.G.D. and G.B.; Data curation: G.B., and A.Ł.L.; Formal Analysis: M.G.D. and G.B.; Funding acquisition: M.G.D.; Investigation: G.B. and M.G.D.; Methodology: M.G.D., G.B. and A.Ł.L.; Software: G.B., A.Ł.L. and M.G.D.; Supervision: M.G.D.; Writing---original draft: G.B. and M.G.D.; Writing---review and editing: M.G.D., G.B., A.Ł.L., and S.C.
All authors have read and agreed to the published version of the manuscript.}

\funding{This research received no external funding.}

\dataavailability{This study uses data supplied by the CDS via anonymous ftp {to} 
\url{http://cdsarc.u-strasbg.fr} (ftp://130.79.128.5) accessed on 13 October 2020 or {via} \url{http://cdsarc.u-strasbg.fr/viz-bin/cat/J/A+A/642/A150} accessed on 13 October 2020.} 

\acknowledgments{G.B. thanks NAOJ because this work started while she was visiting the Division of Science at NAOJ. We thank Matilde Signorini for the fruitful discussion on the physical origin of the intrinsic dispersion of the flux--flux relation. We thank Malgorzata Bogdan for the contribution on the statistics and Takashi Hamana, Guido Risaliti, and Elisabeta Lusso for the discussion on the selection methodology.}

\conflictsofinterest{The authors declare no conflicts of interest.} 

\appendixtitles{yes}
\appendixstart
\appendix
\section[\appendixname~\thesection]{The Impact of Different Binning Approaches on Cosmology}
\label{sec:comparison}

We here detail and discuss how the different binning approaches do not influence the cosmological results.
In particular, we compare the results outlined above and derived from the binning in $\mathrm{log_{10}}(1/(1+\Delta z))$ with the ones obtained with the three different binning approaches detailed in Sections~\ref{sec:widerbins}--\ref{sec:nn}. By examining Table~\ref{tab:bincases}, we can compare results from corresponding cases: the sample in which the untouched sources are not included and the sample in which the untouched sources are added. Based on these samples, we can draw the following conclusions: the values of $\Omega_M$ are always compatible within $0.8 \sigma$ and its uncertainty is between 0.089 and 0.232 in all cases. We can also notice a specific trend: for the samples that include the untouched QSOs, the value of $\Omega_M$ and its uncertainty are smaller compared to the samples in which these sources are discarded. This trend observed for the uncertainty on $\Omega_M$ is valid for all the different selection approaches studied. We note a decrease of the central value of $\Omega_M$, from the sample without untouched sources to the one with untouched sources, in all methodologies except for the case with the binning in $\mathrm{log_{10}}(1/(1+\Delta z))$ (see Table~\ref{tab:bincases}). The reduction in the uncertainty on $\Omega_M$ is due to the different redshift coverage of the two samples (with and without the untouched sources), as already detailed in Section~\ref{sec:goldsampleresults_2}. Considering instead the reduced value of $\Omega_M$, the trend suggests that the untouched sources, mainly located at very low $z$, prefer smaller values of $\Omega_M$. In this regard, we can also comment that the different trend of the case of the binning in $\mathrm{log_{10}}(1/(1+\Delta z))$ could be explained by considering that, compared to the other cases, the number of untouched sources is larger, and they are mainly located not at low redshifts but at $z > 3.5$. Hence, the contribution of these sources is expected to be different from the ones at low $z$, thus not leading to a reduction of $\Omega_M$.
Nevertheless, this topic deserves to be further investigated in future analyses, e.g., \citep{Dainotti2023inprep}. To verify the dependence of high-z QSOs on the precision on $\Omega_M$, we computed $\Omega_M$ with correction for evolution for a whole sample with sources discarded at $z>3.435$ (the maximum z of the 1065 sources sample). The posterior distribution is centered at $\Omega_M\approx 0.39$ with a standard deviation of $0.21$. With only 45 sources discarded from the whole sample, we obtain significantly less precise results for the determination of $\Omega_M$.

In addition, we can also notice from Table~\ref{tab:bincases} that $\Omega_M$ is not constrained in the case of bins centered on each QSO without untouched sources. 
This result clearly shows the crucial role of the compromise between a small intrinsic scatter of the flux--flux relation and the number of sources in view of using QSOs for precision cosmology. Indeed, on the one hand, large samples cannot constrain well $\Omega_M$ due to their large intrinsic scatter. On the other hand, reducing the sample size leads to a reduced intrinsic dispersion and thus to better estimate $\Omega_M$ but only until the number of sources does not become too small to determine $\Omega_M$ with closed contours, as in the case of the 811 QSOs. Moreover, we observe that, when the sample size significantly increases, approaching the initial size of 2421 QSOs, the value of $\Omega_M$ increases toward $\Omega_M {\sim} 0.5$. This completely agrees with the value $\Omega_M= 0.500 \pm 0.210$ reported by~\citep{biasfreeQSO2022} and obtained with the full QSO sample. We here also notice that all the obtained values of $\Omega_{M}$ are compatible with each other in $<1.2\, \sigma$. We show this consistency by reporting in Table~\ref{tab:bincases} the ``$\mathrm{z\text{-}score_{gold}}$'' parameter, which computes the compatibility of each result with the result that presents the smallest uncertainty on $\Omega_M$. The $\mathrm{z\text{-}score_{gold}}$ is computed as $\frac{\Omega_{M,\,i}-\Omega_{M,\, gold}}{\sqrt{\Delta_{\Omega_{M,\,i}}^2+\Delta_{\Omega_{M,\,gold}}^2}}$, where the index $i$ indicates a given value from the table and the index ``gold'' refers to the measurement with the smallest uncertainty: $\Omega_{M,\,gold}=0.229\pm 0.061$. 
Remarkably, the $\Omega_M$ values obtained in this work are also compatible within 1.7$\sigma$ with the most recent measurement of $\Omega_M$ from Pantheon + SNe Ia~\citep{2022ApJ...938..110B}, which is $\Omega_M=0.338 \pm 0.018$. More specifically, our values are compatible with the one reported in~\citep{2022ApJ...938..110B} within 0.9$\sigma$, with the exception of our case of 1125 selected QSOs, which manifests a discrepancy of 1.7$\sigma$. This is shown in the last column of Table~\ref{tab:bincases}, which reports the ``$\mathrm{z\text{-}score_{SNe}}$'' parameter. This parameter is calculated with the same formula of $\mathrm{z\text{-}score_{gold}}$ detailed above, but replacing $\Omega_{M,\,gold}$ with $\Omega_{M,\,SNe} = 0.338 \pm 0.018$. The overall compatibility between the values of $\Omega_M$ obtained with our selected QSO samples and the value of $\Omega_M$ reported in~\citep{2022ApJ...938..110B} significantly proves that the cosmology is reliably recovered independently on the binning approach, and thus on the number of selected sources.  

\section[\appendixname~\thesection]{The Varying Evolution Method}
\label{The varying evolution}

Referring to Section~\ref{epmethod}, we here describe additional details of the 
EP statistical method applied to our selected samples of 1132 and 1065 QSOs considering the simple evolutionary form with $g(z)=1/(1+z)^k$, where $k$ is the slope of the {power-law} (see also Section~\ref{epmethod}). 
This treatment is similar to the application of this correction to the original sample of 2421 QSOs, detailed in~\citep{DainottiQSO}.
{We here stress that we apply the EP method after, and not before, the selection of the QSO sample since we need to apply this correction only once, and we have chosen to apply to the luminosities since anyway the luminosities by definition carry redshift evolution, being dependent on the redshift.}

\begin{figure}[H]
\begin{adjustwidth}{-\extralength}{0cm}
\centering 
\includegraphics[width=8cm]{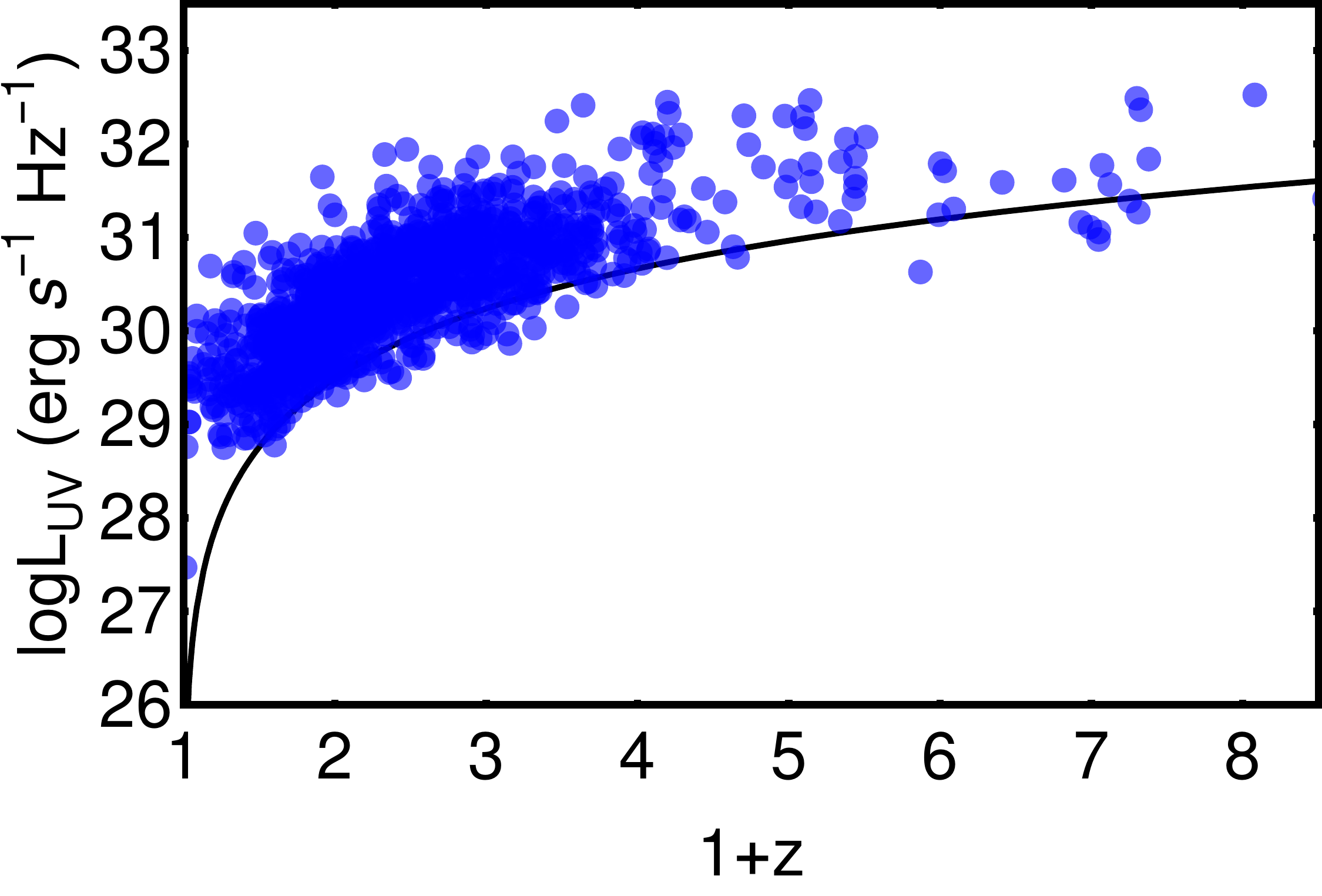}
\includegraphics[width=8cm]{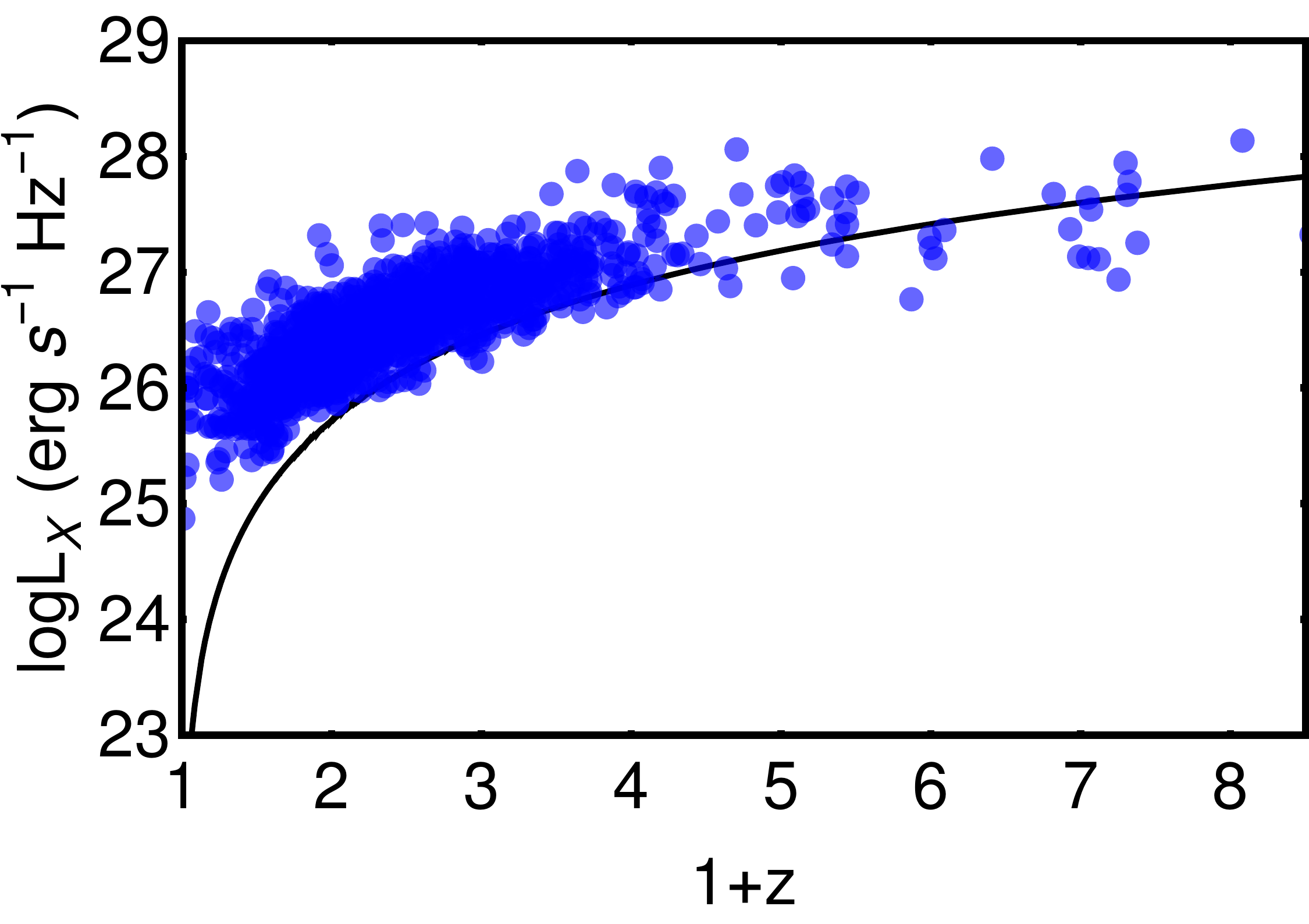}
\end{adjustwidth}
\caption{Redshift evolution of $L_{UV}$ (\textbf{left} panel) and $L_{X}$ (\textbf{right} panel) in units of $\mathrm{erg \, s^{-1}\, Hz^{-1}}$ for the QSO sample of 1132 sources. The black line in both panels shows the limiting luminosity chosen according to the prescription here described.}
\label{fig:evoL-z_1132}
\end{figure}

\begin{figure}[H]
\begin{adjustwidth}{-\extralength}{0cm}
\centering 
\includegraphics[width=8cm]{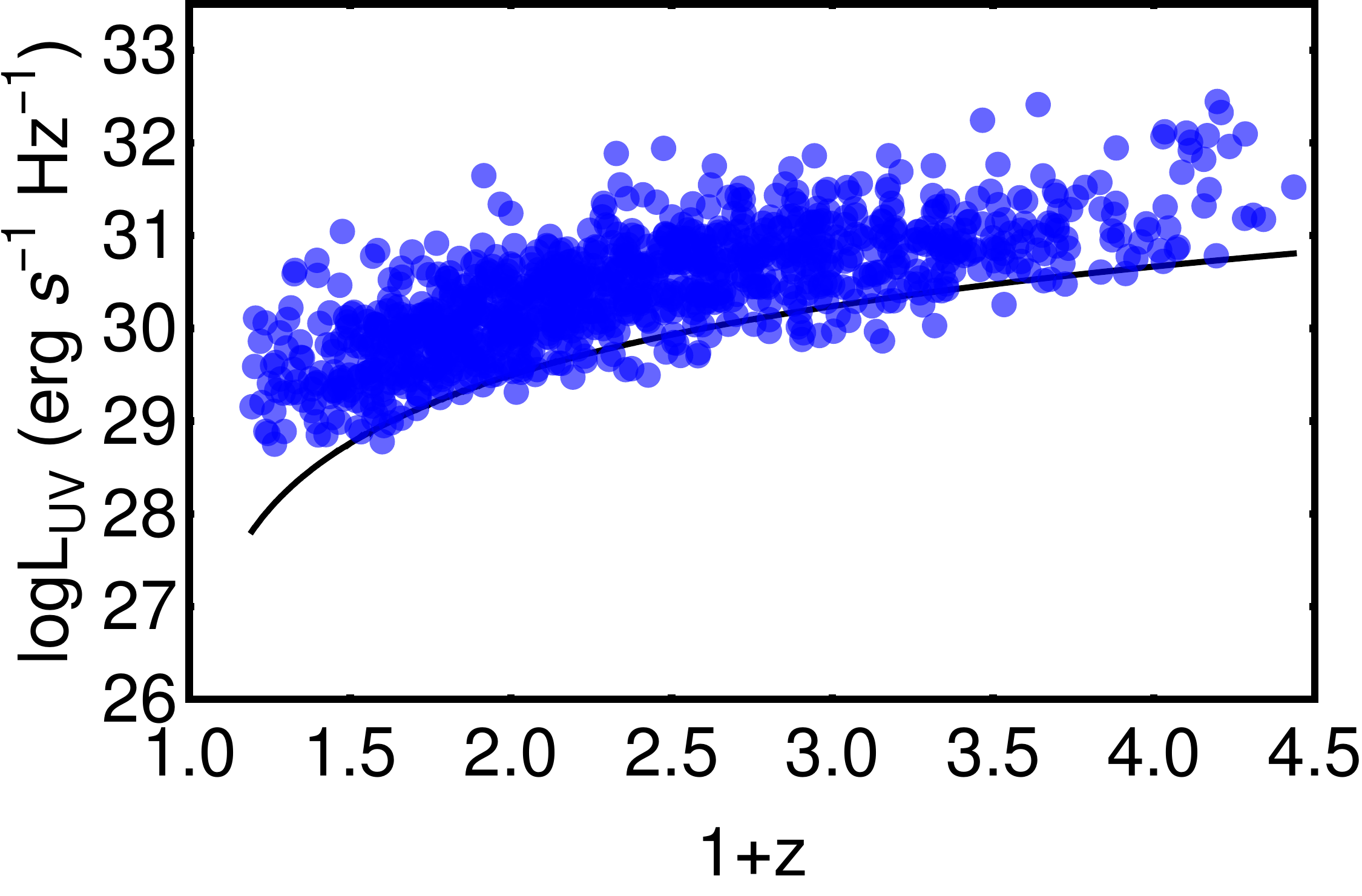}
\includegraphics[width=8cm]{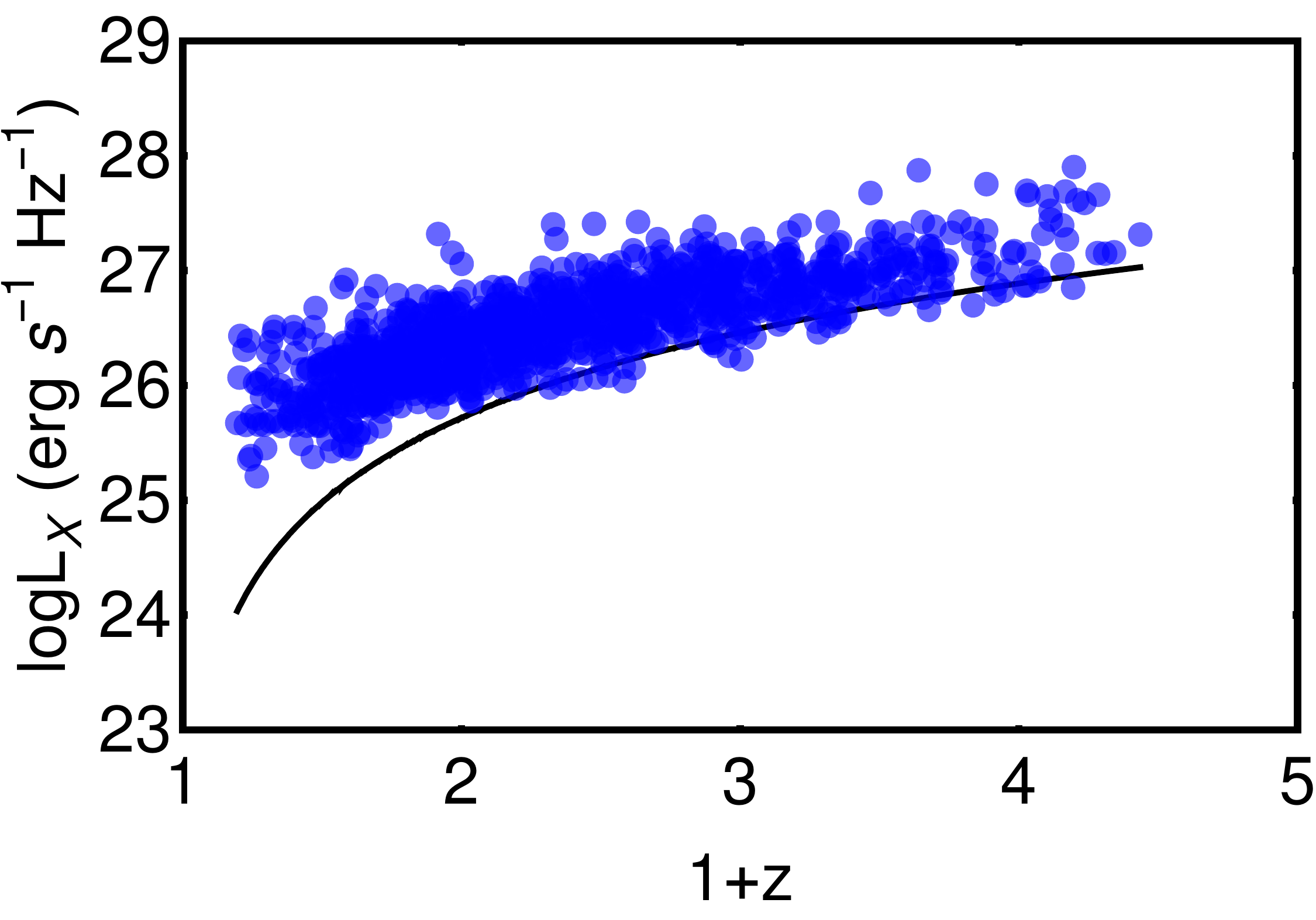}
\end{adjustwidth}
\caption{Redshift evolution of $L_{UV}$ (\textbf{left} panel) and $L_{X}$ (\textbf{right} panel) in units of $\mathrm{erg \, s^{-1}\, Hz^{-1}}$ for the QSO sample of 1065 sources. The black line in both panels shows the limiting luminosity chosen according to the prescription here described.}
\label{fig:evoL-z_1065}
\end{figure}

As anticipated, from the measured flux we compute the luminosity for each QSO assuming a flat $\Lambda \mathrm{CDM}$ model with $\Omega_{M}=0.3$ at the current time and $H_{0} = 70  \, \mathrm{km\,s^{-1}\,Mpc^{-1}}$. 
We also compute the flux limit $F_{lim}$ and the corresponding luminosity $L_{min}(z_i)$. \linebreak  
Specifically, we have chosen \mbox{$F_{lim} = 6\times \, 10^{-29} \,\mathrm{erg \, s^{-1} \, cm^{-2} \, Hz^{-1}}$} for the UV and \linebreak  \mbox{$F_{lim} = 8\times \, 10^{-33}\, \mathrm{erg \, s^{-1} \, cm^{-2} \, Hz^{-1}}$} for the X-rays.
We have also verified through the means of the Kolmogorov--Smirnov (KS) test that, for the full and the cut samples in both X-rays and UV, 
the probability of the null hypothesis that the two samples are drawn by the same distribution cannot be rejected at the \emph{p}-value of $p=0.79$ (for the sample of 1132~QSOs) and $p=0.40$ (for the sample of 1065 QSOs) for the UV and $p=0.50$ (for the sample of 1132~QSOs) and $p=0.38$ (for the sample of 1065 QSOs) for X-rays.
The limiting values for $L_{UV}$ and $L_X$ corresponding to the above-mentioned values of $F_{lim}$ are shown with a black continuous line in the left and right panels of Figures~\ref{fig:evoL-z_1132} and~\ref{fig:evoL-z_1065}, respectively, over the whole set of data points represented by blue filled circles. 

\begin{figure}[H]

\begin{adjustwidth}{-\extralength}{0cm}
\centering
\includegraphics[width=8cm]{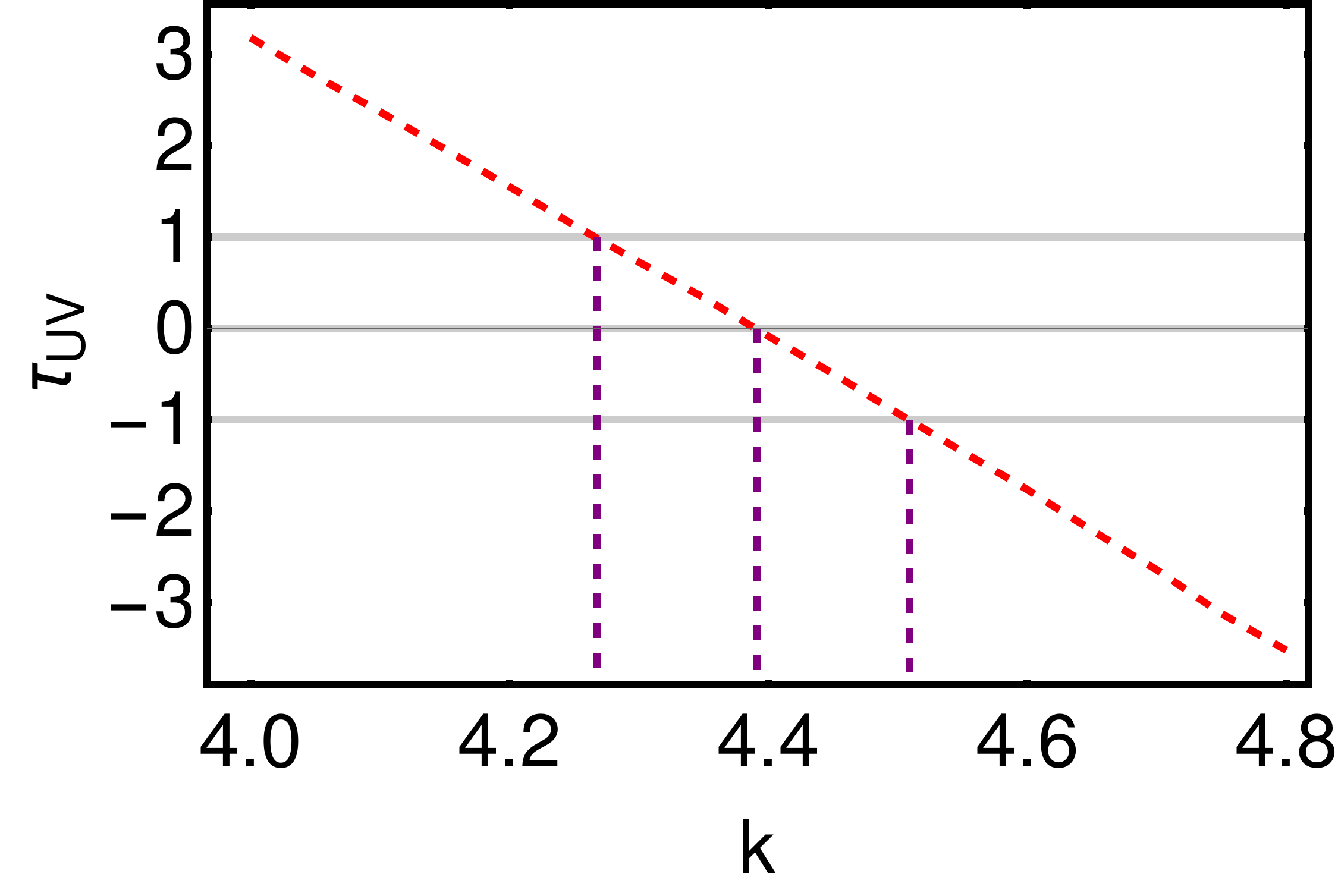}
\includegraphics[width=8cm]{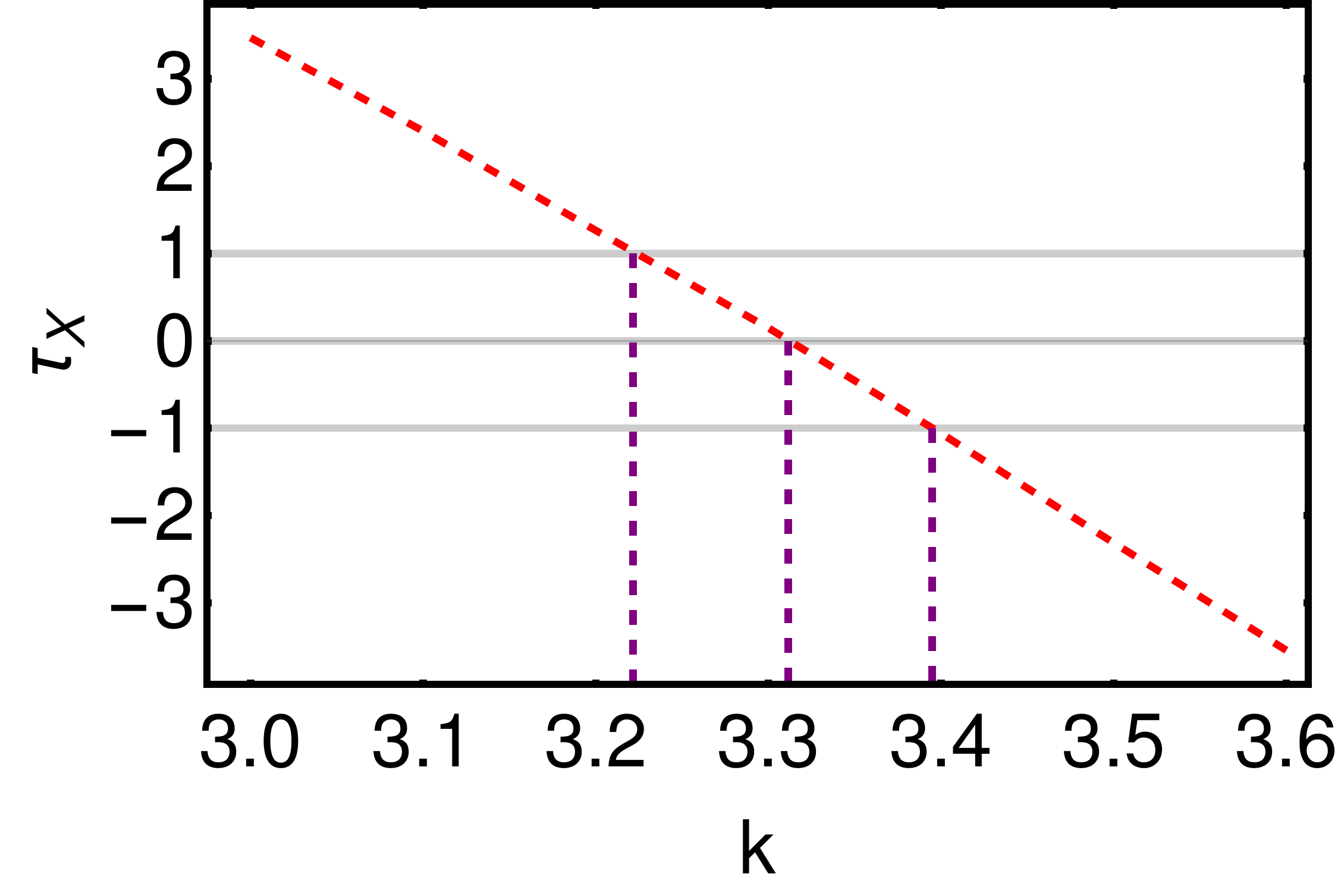}
\end{adjustwidth}
\caption{$\tau (k)$ function (dashed red line) for both the UV (\textbf{left} panel) and X-ray (\textbf{right} panel) analyses for the 1132 QSOs. The point $\tau = 0$ gives us the $k$ parameter for the redshift evolution of $L_{UV}$ and $L_{X}$, while $|\tau| \leq 1$ (gray lines) the 1$\sigma$ uncertainty on it (dashed purple lines). }
\label{fig:evot-k_1132}
\end{figure}

\begin{figure}[H]

\begin{adjustwidth}{-\extralength}{0cm}
\centering
\includegraphics[width=8cm]{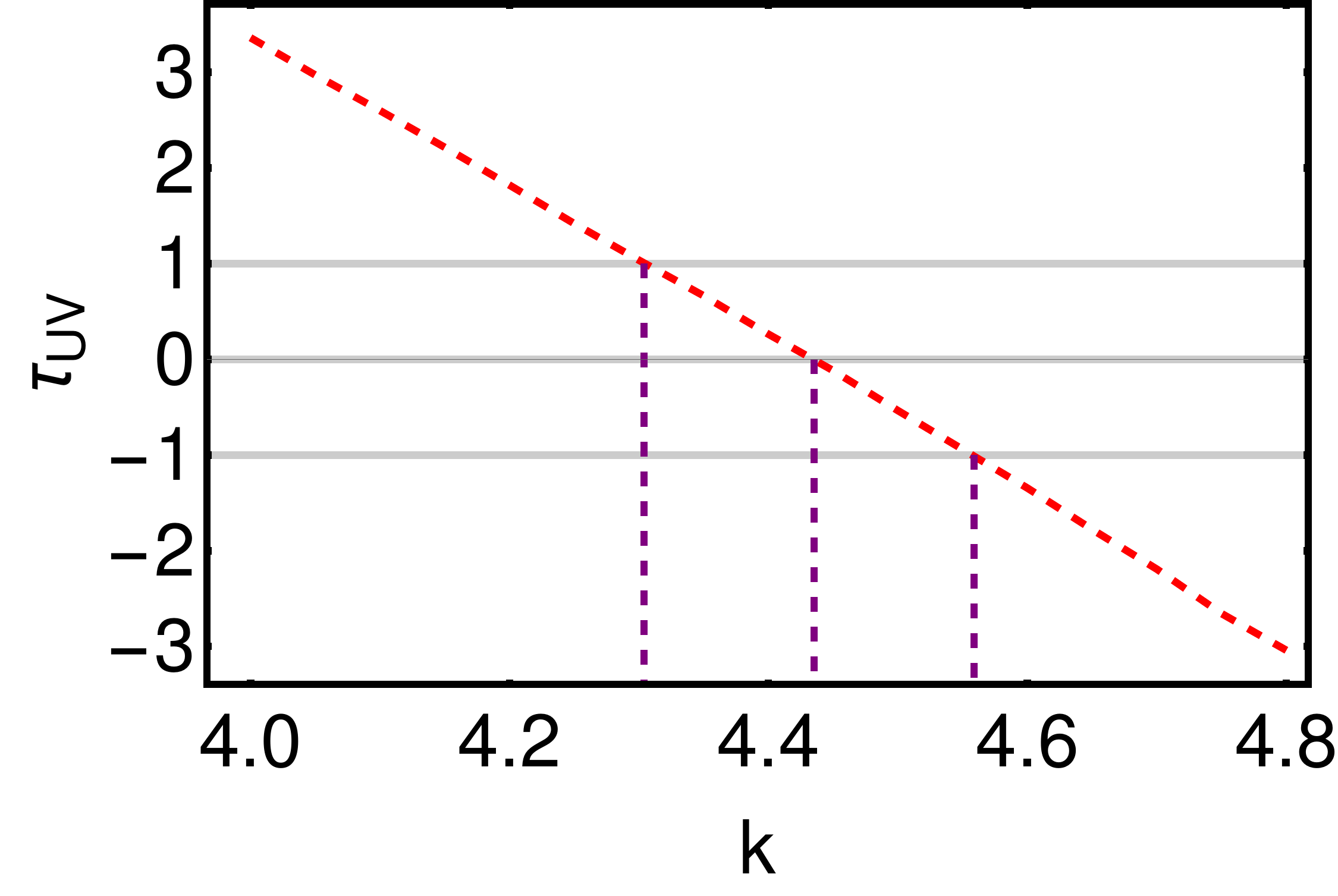}
\includegraphics[width=8cm]{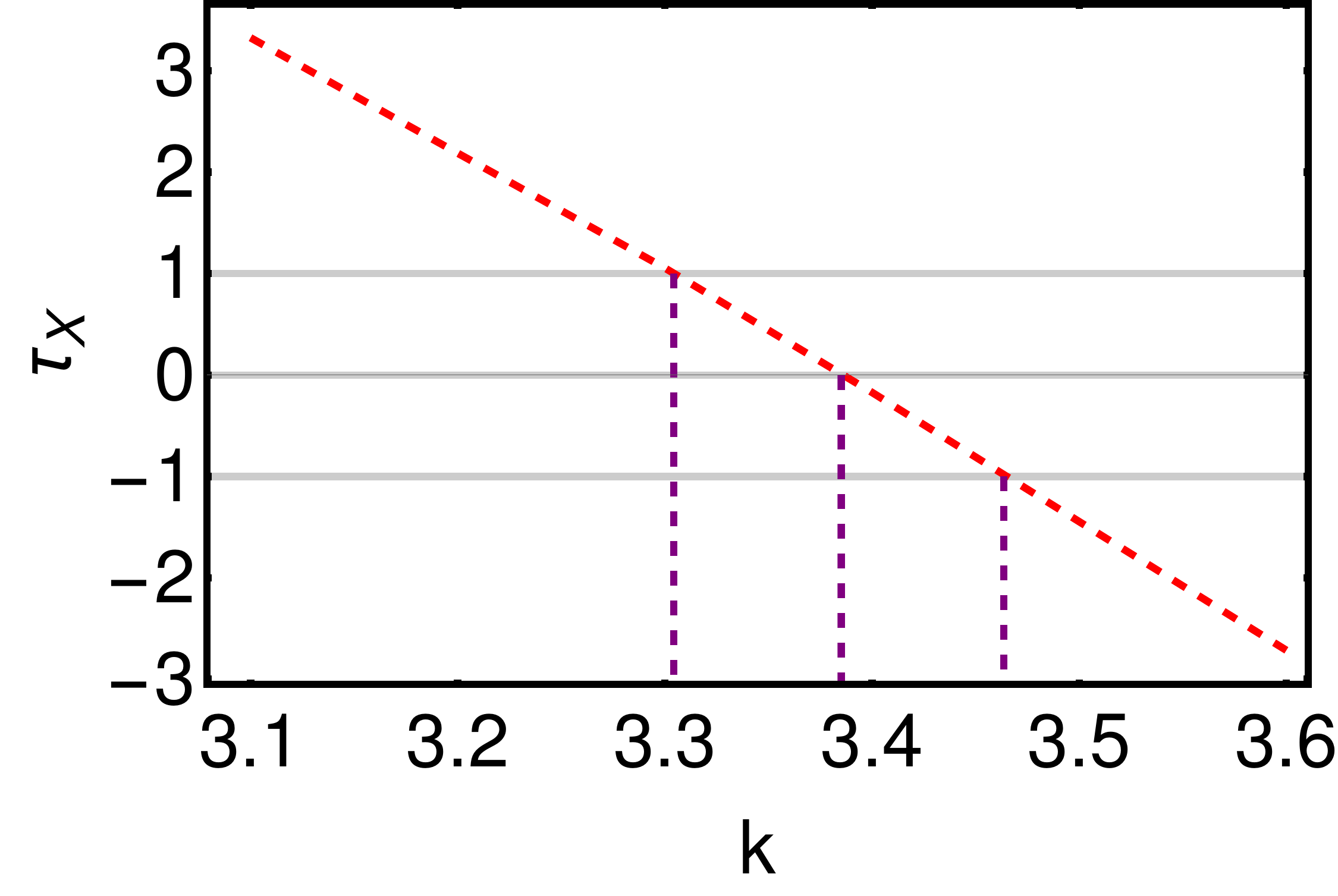}
\end{adjustwidth}
\caption{$\tau (k)$ function (dashed red line) for both the UV (\textbf{left} panel) and X-ray (\textbf{right} panel) analyses for the 1065 QSOs. The point $\tau = 0$ gives us the $k$ parameter for the redshift evolution of $L_{UV}$ and $L_{X}$, while $|\tau| \leq 1$ (gray lines) the 1$\sigma$ uncertainty on it (dashed purple lines). }
\label{fig:evot-k_1065}
\end{figure}

We then apply the $\tau$ test to the data sets trimmed with values of the fluxes mentioned above and obtain the trend for $\tau(k)$ shown in the left and right panels of \mbox{Figures~\ref{fig:evot-k_1132} and~\ref{fig:evot-k_1065}} for the UV and X-rays, respectively. 
For the UV and X-rays, respectively, we obtain $k= 4.39 \pm 0.12$ and $k=3.31 \pm 0.08$ for the sample of 1132 QSOs and $k= 4.44 \pm 0.13$ and $k=3.39 \pm 0.08$ for the sample of 1065 QSOs.
The fact that the values of the evolutionary parameter $k$ are significantly different from 0 for both samples and both in UV and X-ray strongly proves that the redshift evolution plagues our final samples, as well as affects the original sample, as detailed in~\citep{DainottiQSO}.

It is remarkable that the evolutionary function of the UV in our samples is compatible within $2.8$$\sigma$ with the optical evolutionary coefficient $k_{opt}$ obtained in~\citep{2013ApJ...764...43S}, where the same form of $g(z)$ is used. In their paper, they found $k_{opt}=3.0 \pm 0.5$ and corrected the luminosity~function.

\begin{figure}[H]

\begin{adjustwidth}{-\extralength}{0cm}
\centering
\includegraphics[width=8cm]{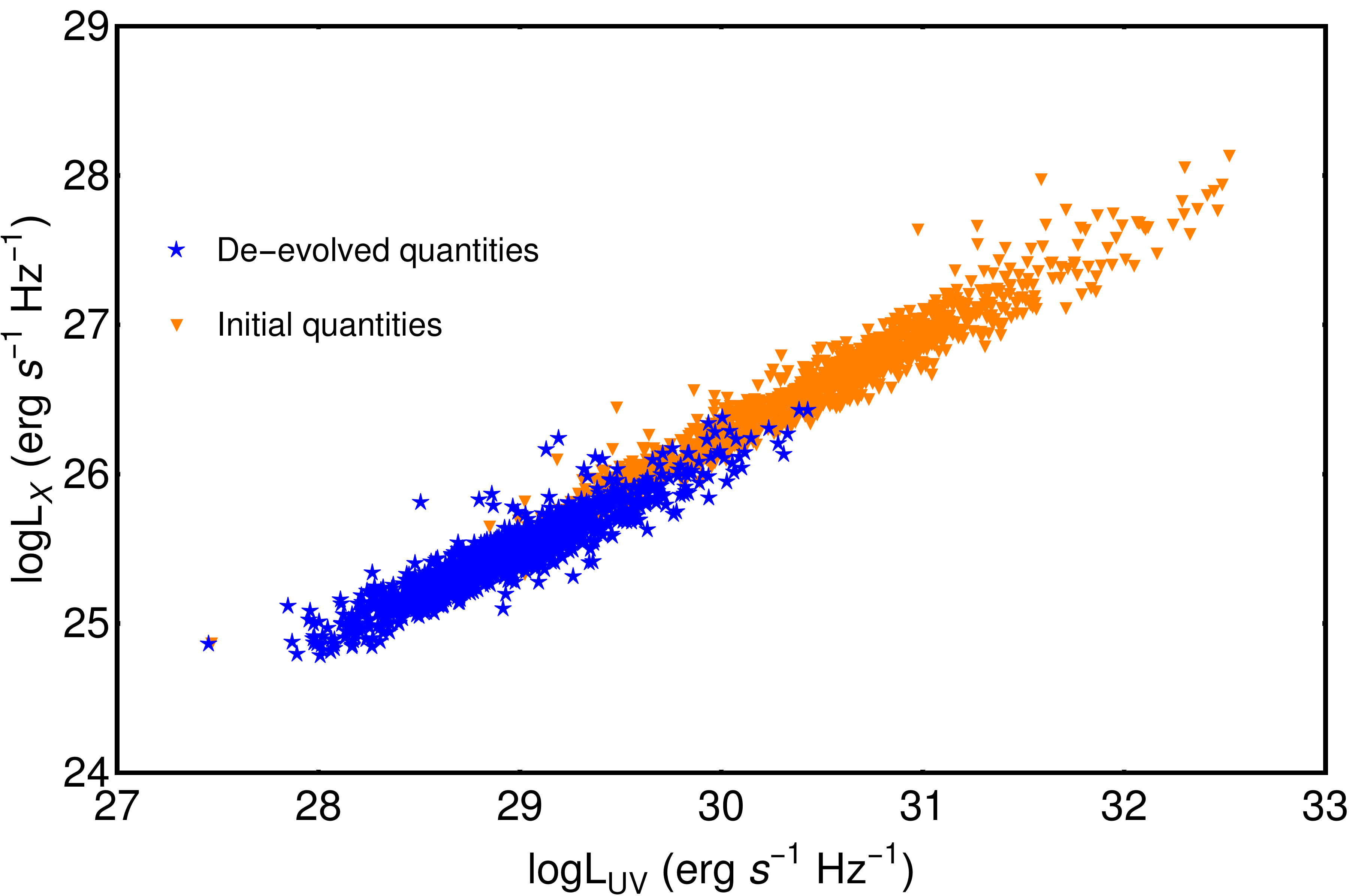}
\includegraphics[width=8cm]{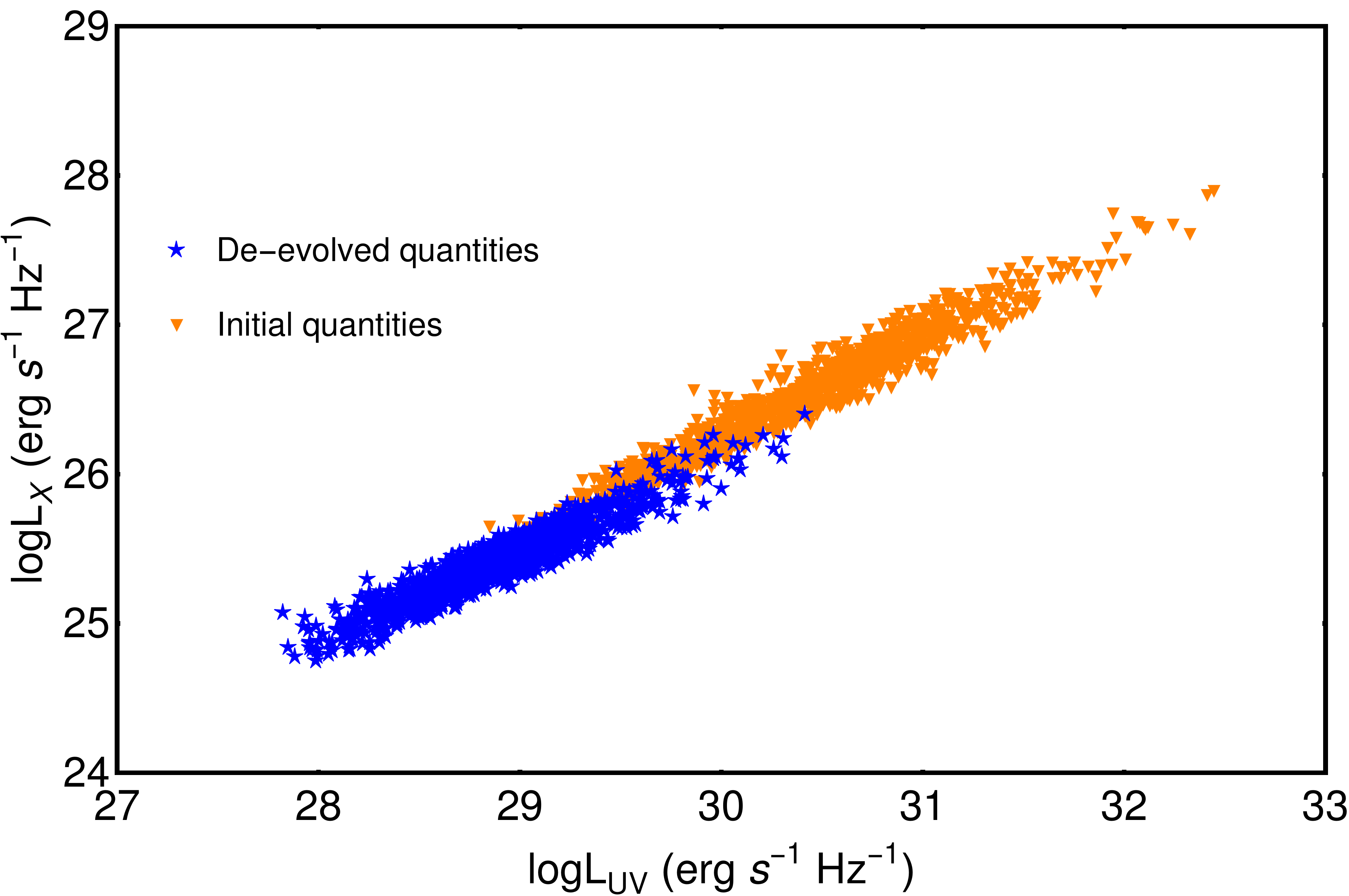}
\end{adjustwidth}
\caption{{Comparison} 
between initial (orange) and de$-$evolved (
blue) quantities in the ($\mathrm{log}L_{UV}$, $\mathrm{log}L_{X}$) plane, respectively, for the 1132 QSOs (\textbf{left} panel) and 1065 QSOs (\textbf{right} panel).}
\label{deev}
\end{figure}

\begin{figure}[H]

\begin{adjustwidth}{-\extralength}{0cm}
\centering
\includegraphics[width=8cm]{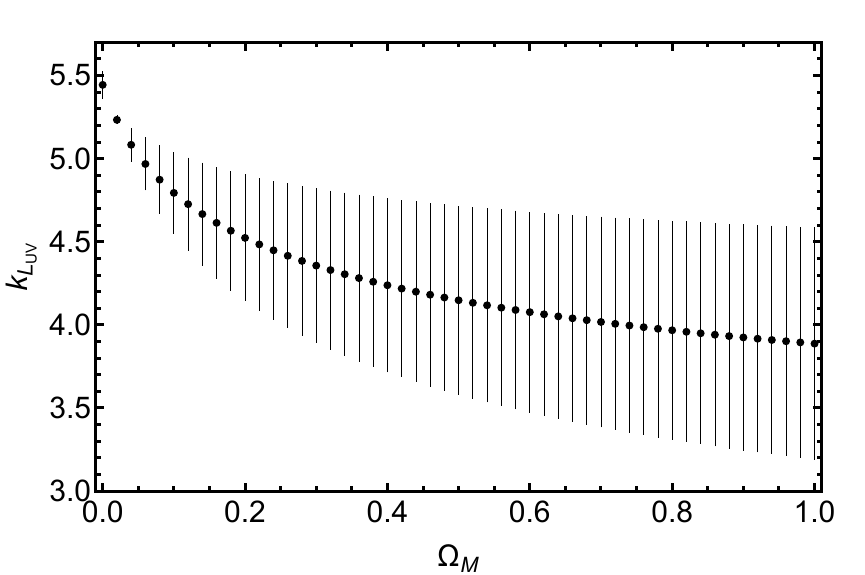}
\includegraphics[width=8cm]{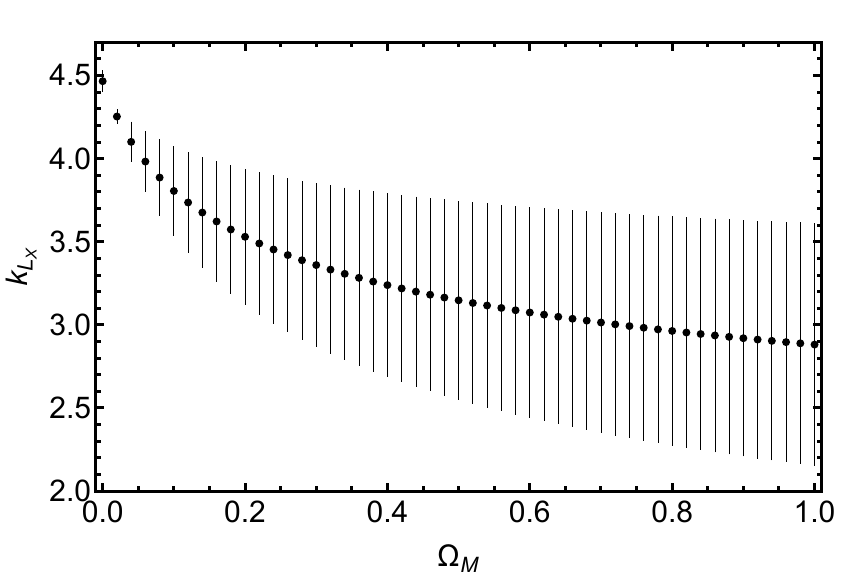}
\end{adjustwidth}
\caption{The results of the EP method obtained for a grid of values of $\Omega_M$ for the sample of 1132 sources. $k_{L_{UV}}(\Omega_M)$ is shown on the \textbf{left}, while $k_{L_{X}}(\Omega_M)$ on the \textbf{right}. The error bars correspond to the 1$\sigma$ confidence intervals.}
\label{fig:evot-k_1131_var}
\end{figure}

After we insert our values of $k$ in $g(z)$, we then compute the new de-evolved luminosities and the associated uncertainties for the whole QSO samples. {The associated uncertainties are computed by propagating the uncertainties of the obtained $k$ values, together with the uncertainties of the initial non-corrected luminosities, in the formula of $L'$ (i.e., $L'= L / (1+z)^k$).} The comparison between these quantities and the initial ones is shown in Figure~\ref{deev} in the ($\mathrm{log}L_{UV}$, $\mathrm{log}L_{X}$) plane for both the sample of 1132 QSOs (left panel) and the sample of 1065 QSOs (right panel). Compared to the initial ones, the computed luminosities span a smaller region of the ($\mathrm{log}L_{UV}$, $\mathrm{log}L_{X}$) plane and show a slightly greater dispersion. This fact is expected because the $g(z)$ function, once the best-fit values for $k$ are used, yields a greater correction (i.e., lower de-evolved values) for higher luminosities. In addition, we have accounted for the error on the determination of $k$ by propagating the errors on the $g(z)$ function. This naturally increases the associated uncertainties on the luminosities. The correction for $g(z)$ affects the spread of the luminosities, hence the dispersion of the correlation, which is consequently larger.
To summarize, the dispersion increases due to the larger spread of the luminosities, and it is minimally affected by the error propagation due to $g(z)$. In other words, the dispersion yielded by the function $g(z)$ is larger than the contribution given by the additional errors due to $g(z)$. Larger errors on the variables may reduce the dispersion, but in this case not sufficiently enough to balance the increase of the dispersion due to the function $g(z)$.
Figure~\ref{deev} clearly shows the effects of the application of the EP method on our data. {Figure~\ref{fig:evot-k_1131_var} shows the values of k obtained for a grid of values of $\Omega_M$ for our gold sample of 1132 QSOs. We here stress that this figure is different from Figure 4 in~\citep{DainottiQSO} in which the full sample of 2421 QSOs is~considered.}

\appendixtitles{no} 

\begin{adjustwidth}{-\extralength}{0cm}

\reftitle{References}




\PublishersNote{}
\end{adjustwidth}
\end{document}